\documentclass[aip,amsmath,amssymb,reprint]{revtex4-1}

\usepackage{graphicx}    
\usepackage{dcolumn}
\usepackage{bm}

\usepackage[utf8]{inputenc}
\usepackage[T1]{fontenc}
\usepackage{mathptmx}
\usepackage{enumitem}

\newcommand{\ud}{\mathrm{d}}

\newcommand{\F}{{\bf F}}
\newcommand{\B}{{\bf B}}
\newcommand{\J}{{\bf j}}

\newcommand{\E}{{\bf E}}
\newcommand{\V}{{\bm v}}

\newcommand{\K}{{\bm k}}

\newcommand{\RR}{{\bf R}}
\newcommand{\parent}[1]{\left(#1\right)}
\newcommand{\abrack}[1]{\left[#1\right]}
\newcommand{\cbrack}[1]{\left\{#1\right\}}

\newcommand{\dive}{\nabla\cdot}

\newcommand{\dotg}{\cdot\nabla}

\newcommand{\ez}{\hat{\bf z}}

\newcommand{\rhat}{\hat{\bf r}}
\newcommand{\that}{\hat{\bm{\theta}}}
\newcommand{\zhat}{\hat{\bf z}}
\newcommand{\xhat}{\hat{\bf x}}
\newcommand{\yhat}{\hat{\bf y}}

\newcommand{\sci}[2]{#1\times10^{#2}}

\newcommand{\refeq}[1]{(\ref{eq:#1})}
\newcommand{\Ommax}[1]{\Omega_{E,\mathrm{Bri}}^{(#1)}}
\newcommand{\omnorm}{\omega_{cp}}

\begin{document}

\preprint{AIP/123-QED}

\title[Two-species Ohkawa filter]{Impact of azimuthal forcing on the Brillouin limit in a collisional two-species Ohkawa filter}

\author{T. Nicolas}
\email{timothee.nicolas@polytechnique.edu.}
\affiliation{CPHT, Ecole Polytechnique, CNRS, 91128 Palaiseau, France}

\date{\today}

\begin{abstract}
  This paper investigates the physics of plasma separation in a two species rotating collisional Ohkawa filter, when the source of rotation is an orbital angular momentum carrying wave. The electric field is treated self-consistently with ion and electron radial motion. The injection of angular momentum causes radial currents leading to charge penetration and electric field build up. The electric field varies until an equilibrium with the friction forces is reached. Both collisions with neutrals and Coulomb collisions are considered. In the case where the electric field is driven by the resonant wave, there is no collisional breakdown of the Brillouin limit~\cite{raxBreakdownBrillouinLimit2015}, on the contrary the maximum achievable electric field decreases when the collision frequency is increased. When two species are present, one that undergoes the wave forcing while the second does not interact with the wave, we find the following: the first species is confined, while the second species can be expelled or confined depending on the charge to mass ratio and the collisionalities. Assuming equal charge numbers, if the second species is the heavy one, it is always expelled, which is a standard result. When the second species is the light one, it can also be expelled in the common case where neutral collisions dominate over Coulomb collisions, which constitutes a new result.
\end{abstract}

\maketitle

\section{\label{sec:intro}Introduction}

Plasma separation, or plasma mass filtering, is the process by which two initially mixed species are divided into two different streams within a plasma, thus separating them. The potential applications include recycling of alloys (\emph{e.g.} rare earth metals) into pure elements~\cite{gueroultOpportunitiesPlasmaSeparation2018}, or efficient treatment of nuclear waste~\cite{sicilianoPlasmaTechniquesReprocessing1993,ahlfeldApplicationDesignProject2005,voronaPossibilityReprocessingSpent2015,gueroultPlasmaFilteringTechniques2015,amirovStudyFeasibilityDistributed2015,dolgolenkoSeparationMixturesChemical2017,smirnovFeasibilityPlasmaSeparation2019}. Regarding the latter, the promise of plasma separation is that of efficiently isolating the small quantity of very dangerous actinide waste from the rest of the low activity waste (structural waste as well as the other products of fission, \emph{e.g.} lanthanides). The advantage over currently used chemical techniques is that the separation is done in one single pass in the plasma device, without further contamination of fluid effluents~\cite{sicilianoPlasmaTechniquesReprocessing1993}. Another merit of plasma separation is that it takes advantage of a significant difference in mass, where chemical separation is made difficult by chemical similarities between actinides and lanthanides~\cite{petermanSeparationMinorActinides2008}. 

Although no plasma separation process is routinely used as of today, it has long been known that plasmas may be tailored for such a purpose\cite{lehnertPartiallyIonizedPlasma1970}. The particles in the plasma are charged, so that they may have different motion according to their mass over charge ratio in a given electromagnetic configuration. The essential difference with a mere spectrometer, which makes plasmas interesting for industrial, rather than diagnostic, uses, is that plasmas are usually close to quasineutrality. This allows to circumvent the insurmountable energy cost that would otherwise be associated with the charge separation, and in principle permits high flux operation at an economical cost, which is required to make industrial applications viable. 

Several different concepts exist for plasma separation (see Ref~\cite{zwebenPlasmaMassSeparation2018} for a recent review), owing to the diversity of imaginable electromagnetic configurations, only limited by Maxwell's equations. In this paper, we shall discuss the Ohkawa filter configuration\cite{ohkawaBandGapIon2002,shinoharaInitialTrialPlasma2007}, in which a radial electric field within an axially magnetized plasma column can cause radial ejection of charged particles, or not, depending on their mass/charge ratio. 
The central concept behind the Ohkawa filter is the Brillouin limit, 
a term that originally pertained to non-neutral plasmas~\cite{brillouinTheoremLarmorIts1945,davidson1974theory,theissRigidRotorEquilibriaNonneutral1975}. A rotating magnetic plasma with charge and mass $(q,m)$ can be radially confined only if the electric field $E$ does not exceed the Brillouin limit, $|E/r|<qB_0^2/(4m)$, where $r$ is the radial coordinate of the plasma column and $B_0$ the homogeneous static magnetic field. Since the Brillouin limit depends on the mass to charge ratio, we can have in principle one population confined while the other would be radially ejected. 

Experimentally, it is possible to build up electric field in a cylindrical plasma column using sets of differentially biased circular electrodes at the ends of the cylinder~\cite{gueroultInitialExperimentalTest2016}. The advantage is that one can have in principle a lot of flexibility for external control, by simply modifying the bias. The problem is that there is no guarantee that the potential will propagate along the magnetic field lines to the plasma~\cite{gueroultNecessaryConditionPerpendicular2019}. Another important historical problem with the electrodes is that of the critical ionization velocity~\cite{laiReviewCriticalIonization2001}. There is, however, another way of generating the electric field, that does not suffer from the aforementioned two problems. It is based on the injection of angular momentum by a wave onto ions~\cite{fettermanChannelingRotatingPlasma2008,fettermanWaveDrivenRotationSupersonically2010}. From a wave-particle interaction point of view, the resonant hamiltonian interaction between the wave and the particle results in a radial displacement of the ion, creating charge separation~\cite{raxEfficiencyWavedrivenRigid2017}. From a fluid point of view, the azimuthal momentum input on the ions results in radial flows. The associated divergent currents are the source of the charge separation, and an equilibrium establishes when the inertial, electric, Lorentz and friction forces balance. 
An advantage of the wave method above the one using electrodes is that the interaction between the walls and the outflow of separated material is reduced~\cite{fettermanWaveDrivenRotationSupersonically2010}. However, there are also significant physical differences in the behaviour of the plasma with respect to the Brillouin limit. The purpose of the present paper is to highlight these differences.

When the electric field is generated through electrodes, one can assume the electric field to be an external control parameter. Of course, since the radial electric field has a non-vanishing divergence, a plasma is necessary in between the electrodes. But to a certain extent, one can consider to fix the electric field arbitrarily by merely turning the knob of the potentiometer. In this context, it has been shown that friction with the neutrals cause a breakdown of the Brillouin limit~\cite{raxBreakdownBrillouinLimit2015}. The transition between the regime where the plasma is confined and that where it is expelled (the Brillouin limit) becomes smooth. When the electric field instead results from the radial balance between inertial, electric and Lorentz forces, under the azimuthal forcing introduced by the wave coupling to the plasma, the results obtained in ref.~\cite{raxBreakdownBrillouinLimit2015} no longer hold. On the contrary, we show that the maximum achievable electric field decreases when the collision frequency is increased. Another important difference is that the expelled species does not have to be the heavy one. Instead, depending on the collisionality, it can be the light one. We show that basically, when friction with neutrals dominates over Coulombian friction (a common situation~\cite{gueroultNecessaryConditionPerpendicular2019}), the confined species is the one that undergoes the forcing, while the other is expelled.


The paper is organized as follows. The framework for the present theoretical study is introduced in section~\ref{sec:Model}. The behaviour of the Brillouin limit in a collisional plasma forced azimuthally is explained in section~\ref{sec:CollisionalBrillouin}. In section~\ref{sec:BrillouinBreakdown}, we focus on the interpretation of the collisional modification of the Brillouin limit in the present context, as compared to that of refs.~\cite{raxNonlinearOhmicDissipation2019,raxBreakdownBrillouinLimit2015}. We consider the behaviour of a trace passive impurity in a plasma where the dominant species is azimuthally forced in section~\ref{sec:trace}, and then the behaviour of the plasma when the two species have comparable abundance in section~\ref{sec:notrace}. We discuss and summarize the results respectively in sections~\ref{sec:Discussion} and~\ref{sec:Summary}.  

\section{\label{sec:Model}Theoretical framework}

\subsection{\label{subsec:Notations}Notations}
We use the usual cylindrical unit vector basis $(\rhat,\that,\zhat)$. The plasma is immersed in a static homogeneous axial magnetic field $\B=B_0\ez$. We do not consider its modifications, which is justified if the ion skin depth is larger than the radius $a$ of the plasma column\cite{gueroultTendencyRotatingElectron2013}. In the numerical applications, a normalization is employed. The radius $a$ serves to normalize the distances. The time and frequencies are normalized using the proton cyclotron frequency $\omnorm\equiv eB_0/m_p$. The velocities are naturally normalized to $a\omega_{cp}$. We will assume for the electric field a form $\E = rE_0/a\rhat$, and we will describe it using the frequency $\Omega_E = E_0/(aB_0)$. $\Omega_E$ can be positive (electric field directed outward) or negative (directed inward), but in the case of the Ohkawa filter, which relies on the Brillouin limit, the relevant sign is the positive one, as will soon be clear. Other separation schemes rely on an electric field pointing inward, see for instance refs.~\cite{bardakovAdvancesProblemsPlasmaoptical2014,liziakinPlasmaMassSeparation2021}. 
The ion (charge numbers $Z_i$ and $A_i$) and electron cyclotron frequencies are denoted respectively $\omega_{ci} = \frac{Z_i}{A_i}\omnorm$ and $\omega_{ce} = \frac{m_p}{m_e}\omnorm$.
Note the sign convention, $\omega_{ce}>0$. The angular frequency of species $i$ is denoted $\Omega_i$. Finally, we introduce the notation
\begin{align}
  \label{eq:ommax}
  \Ommax{i} = \frac{Z_i}{4A_i}\omnorm =\frac{\omega_{ci}}{4},
\end{align}
which is the Brillouin limit for a species with charge and mass numbers $Z_i$ and $A_i$.

\subsection{\label{subsec:fluid}The Brillouin limit}

The Brillouin limit~\cite{davidson1974theory,raxBreakdownBrillouinLimit2015,gueroultCentrifugalInstabilityRegime2017} appears when one is interested in the rigid rotor solution of the ion momentum balance equation in a collisionless plasma without pressure gradient:
\begin{align}
  \label{eq:ionmom_vec}
  \V_i\dotg\V_i = \frac{q}{m}\parent{\E+\V_i\times\B}
\end{align}

If one assumes rigid rotor, then the velocity is written $\V_i = \upsilon_r(r)\rhat + r\Omega_i\that$, with $\Omega_i$ independent of $r$. Projecting along $\rhat$ and $\that$ yields,
\begin{align}
  \label{eq:Brillouinrad}
  &\upsilon_r\upsilon_r' - r\Omega_i^2 = r\omega_{ci}\parent{\Omega_E + \Omega_i}\\
  \label{eq:Brillouinpol}
  &\parent{2\Omega_i + \omega_{ci}}\upsilon_r= 0,
\end{align}
where prime denotes radial derivation. Unless $\Omega_i = -\omega_{ci}/2$, the solution must have vanishing radial flow, and equation \refeq{Brillouinrad} then gives the relation between $\Omega_E$ and $\Omega_i$. The solutions are
\begin{align}
  \label{eq:OmegaBrillouin}
  \frac{\Omega_i}{\omega_{ci}} = -\frac{1}{2}\parent{1 \pm\sqrt{1-4\frac{\Omega_E}{\omega_{ci}}}},
\end{align}
which develops, for small $\Omega_E$, into the so-called slow mode at $\Omega_i\sim -\Omega_E$, and fast mode, at $\Omega_i\sim -\omega_{ci}+\Omega_E$. If $\Omega_E$ is larger than $\omega_{ci}/4$, formally, a solution to \refeq{Brillouinrad}-\refeq{Brillouinpol} still exists if we allow for $\upsilon_r\ne 0$. Assuming the radial velocity to be of the form $\upsilon_r=r\gamma_i$ (we will see in section \ref{subsec:waveforcing} under what conditions this assumption is justified), then the system \refeq{Brillouinrad}-\refeq{Brillouinpol} reads
\begin{align}
  \label{eq:Brillouinradrigid}
  &\Omega_i^2 - \gamma_i^2 + \omega_{ci}\parent{\Omega_E+\Omega_i}=0\\
  \label{eq:Brillouinpolrigid}
  &\parent{2\Omega_i+\omega_{ci}}\gamma_i = 0.
\end{align}
The solution for $\Omega_E>\omega_{ci}/4$ is:
\begin{align}
  \label{eq:brokedownOm}
  &\frac{\Omega_i}{\omega_{ci}} = -\frac{1}{2}\\
  \label{eq:brokedownGam}
  &\frac{\gamma_i}{\omega_{ci}} = \pm\frac{1}{2}\sqrt{-1+4\frac{\Omega_E}{\omega_{ci}}}.
\end{align}

However, there is a problem with this solution. Since the radial velocity of the ions no longer vanishes, there is an inflow or outflow of ions, which modifies the electric field. For instance in the case of the plus sign (outflow), ions are depleted, leaving place to only electrons, and the sign of the electric field can not even remain positive (pointing outwards). The Brillouin limit raises many questions in the context of the Ohkawa filter. If the principle is to expel radially the heavy fraction of a two-species plasma by overcoming the Brillouin limit, but the Brillouin limit cannot be overcome, what remains of the Ohkawa filter? Is there actually a way to charge the plasma at $\Omega_E>\omega_{ci}/4$, and how does the plasma behave when there are two species? Incidentally, how is the electric field generated in the first place? The purpose of the following sections is to give clear answers to these interrogations, in the context where the source of angular momentum is an orbital angular momentum carrying wave coupled to the plasma.

\subsection{\label{subsec:waveforcing}Wave forcing and ion momentum equation}
The most important assumption of this paper is to never separate the electric field from the source that causes it. Since we will consider collisional friction, there cannot be a solution with finite rotation unless we explicitly take into account the external source of momentum and energy. We will assume this source to be a wave resonant with an ion species. The resonance is at a frequency related to the cyclotron resonance, although the specific resonance condition also depends on the axial wave number and the electric drift rotation frequency~\cite{fettermanChannelingRotatingPlasma2008}. We will introduce another ion species, not affected by the wave. We will refer to the former as the forced species, and to the latter as the passive species. Throughout this work, the ``first species'' will always be the forced one, denoted by an index 1, while the ``second species'' will be the passive one, denoted by an index 2. Each quantum of a wave with spatiotemporal dependence $\exp\parent{i(kz+l\theta - \omega t)}$ carries energy $\hbar\omega$, linear momentum $\hbar k\zhat$ and angular momentum $\hbar l\zhat$. If it can be coupled resonantly to the ions, then the resonant ions are displaced radially. The associated radial current is called quasilinear current, because the theory that describes the resonant absorption of the wave is quasilinear theory. However, if one assumes the absorbed power is known, it is legitimate to model the interaction in a fluid model with a force acting on the ions, because the momentum content of the wave per unit energy is fixed by its frequency and wave numbers. The value of the force density is ${\bf F} = \K P/\omega$, where $P$ is the power density absorbed, $\omega$ the wave frequency and $\K$ its wave vector. This approach is all the more legitimate because we are interested in the component of the force that is in the perpendicular direction. In the parallel direction, there is also a net transfer of momentum equal to $\hbar k_z$ times the number of quanta absorbed per unit time ($F_\| = k_\|P/\omega$), but this momentum is passed to the bulk ions on a collisional time scale only. In the perpendicular direction, the momentum is passed to the bulk ions almost instantaneously, on the cyclotron timescale ~\cite{guanPlasmaRotationInduced2013,guanToroidalPlasmaRotations2013} (see also appendix~\ref{ap:justif}). In this paper, we neglect the force in the parallel direction.

The modeling of the plasma wave interaction in terms of a fluid force needs to be thoroughly justified, in particular because it is different from what was done in previous works. For instance, in ref.~\cite{raxBreakdownBrillouinLimit2015}, the effect of the wave is completely absent from the fluid equations. The latter are used to derive the collisional fluxes, essentially the nonlinear current inducing the charge relaxation, and then it is merely said that the generator compensates by replenishing the charge:
\emph{``The collisional depletion of $\rho$ through $\J$ is continuously compensated by the power supply driving these electrodes, or through wave induced charge separation, in order to ensure steady state rotation.''}
But the fluid model used in this reference, which does not explicitly include the effect of the wave, cannot lead to a steady state. It is easy to understand why: there is a collisional term of angular momentum loss (the neutrals at rest), but the equations don't contain any term to compensate for this loss. It is, indeed, the wave induced current (or current driven through the electrodes in the case of electrode driving), but it does not appear in the equations. As a result, the steady state that is obtained is inconsistent. Another way to put it is that there is no evolution equation for the electric field. In contrast, the present approach takes into account explicitly not only the resonant current induced by the wave, but also the polarization currents that are essential to set the bulk plasma into motion. However, the treatment of the wave resonant interaction inducing a force density ${\bf F} = \K P/\omega$, is not new. A similar line of thought is adopted in refs.~\cite{guanPlasmaRotationInduced2013,guanToroidalPlasmaRotations2013}, in the context of toroidal rotation in tokamaks as a side effect of current drive with lower hybrid wave. Since the momentum of $\E\times\B$ motion is carried mainly by the ions, while the resonant particles are electrons in this context, they don't include the force in the ion fluid equation, and rather interpret it as a force on the electric field. Nonetheless, the present approach is clearly inspired from these references. In appendix~\ref{ap:justif}, we give more technical justifications in a slab configuration (which allows to neglect inertial effects) equivalent to a capacitor where the dielectric is a magnetized plasma.

The question of the characterization of the wave, the conditions for its propagation and resonant deposition on the ions, which pertain to quasilinear theory, are beyond the scope of this paper. See refs~\cite{fettermanWaveParticleInteractions2011,raxEfficiencyWavedrivenRigid2017} for some insight in this important matter. We will simply assume that at $t=0$ in a magnetized plasma column initially at rest, the ions are set into azimuthal motion by a force representing the action of this wave. We will assume that power deposition can be tailored so as to follow a law $\mathcal{P}(r)\sim r^4$, where $\mathcal{P}(r)$ represents the total power per unit axial length deposited within a surface delimited by the radius $r$. Put differently, the local power per unit volume deposited by the wave must be proportional to the square of the radius. Setting aside the potential practical difficulty, this has the immense advantage that the complicated set of partial differential equations can be reduced to ordinary differential equations (ODEs) in that case. To see why, consider a shell of infinitesimal thickness $\ud r$, with mass per unit axial length $\ud m = 2\pi \rho r\ud r$, $\rho$ being the mass density of the plasma. In this shell, a number $\ud \dot{N}$ of wave quanta per unit time are absorbed by the ions, with $\hbar\omega\int_0^r\ud\dot{N} = \mathcal{P}(r)$. Writing the angular velocity as $r\Omega(r)$, where \emph{a priori} $\Omega$ depends on $r$, the angular momentum theorem (considering the wave forcing as the only force) writes
\begin{align}
  \label{eq:angmomth}
  2\pi r^3 \dot{\Omega}\rho \ud r = \hbar l\ud \dot{N}.
\end{align}
Therefore, $\dot{\Omega}$ is independent of $r$ if $\ud\dot{N}\sim r^3\ud r$, meaning $\mathcal{P}(r)\sim r^4$. We neglect the transfer of linear momentum, as well as the wave heating. We assume a homogeneous density for all species in the plasma column, and take into account the friction between charged species, and with neutrals at rest having density $n_0$. The ion momentum equation is therefore modeled as
\begin{eqnarray}
  \label{eq:ionmomeq}
  \parent{\partial_t\V_1 + \V_1\dotg\V_1} &=& \frac{Z_1e}{m_1}\parent{\E+\V_1\times\B}+ \frac{{\bf F}_\mathrm{wave}}{m_1n_1}\nonumber\\
  && - \sum_{s\ne 1}\nu_{1s}\parent{\V_1-\V_s} - \nu_{10}\V_1,
\end{eqnarray}
where $\nu_{ss'}, s\geq 1$ is a Coulomb collision frequency, $\nu_{s0}, s\geq 1$ is a frictional collision frequency with the neutrals and ${\bf F}_\mathrm{wave}$ represents the wave-induced torque:
\begin{align}
  {\bf F}_\mathrm{wave} = \frac{1}{2\pi r^2}\frac{\ud\mathcal{P}}{\ud r}\frac{l}{\omega}\that\propto r\that
\end{align}
Because of the radial variation of ${\bf F}_\mathrm{wave}$, we can assume the azimuthal velocities to be of the form $\upsilon_\theta = r\Omega$, where $\Omega$ does not depend on $r$. Because of the rigid body rotation allowed by the choice of power deposition, the assumption that $\E = E_0 r/a\rhat$, corresponding to $\Omega_E$ also independent of $r$, is legitimate as well. Since such an electric field can only arise from the divergence of radial currents, this means that the divergence of these currents also does not depend on $r$. Indeed $\dive\E = 2E_0/a$. This means that the radial velocities can legitimately also be taken of the form $\upsilon_r = r\gamma$, where $\gamma$ does not depend on $r$. The assumptions that ${\bf F}_\mathrm{wave}\sim r\that$ and $\V = r(\gamma\rhat + \Omega\that)$ are what allows to transform the PDEs into ODEs.

In projection along $\rhat$ and $\that$, equation \refeq{ionmomeq} yields the set of equations:
\begin{align}
  \label{eq:gammadot}
  &\dot{\gamma}_1 = \Omega_1^2 - \gamma_1^2 + \omega_{c1}\parent{\Omega_E+\Omega_1} - \sum_{s\ne 1}\nu_{1s}\parent{\gamma_1 - \gamma_s} - \nu_{i0}\gamma_1\\
  \label{eq:Omegadot}
  &\dot{\Omega}_1 = -\parent{2\Omega_1+\omega_{c1}}\gamma_1 + \omnorm^2F  - \sum_{s\ne 1}\nu_{1s}\parent{\Omega_1 - \Omega_s} - \nu_{i0}\Omega_1,
\end{align}
where
\begin{align}
  F = \frac{1}{2\pi \rho_1 r^3\omnorm^2}\frac{\ud\mathcal{P}}{\ud r}\frac{l}{\omega} < 0
\end{align}
We choose the forcing to be negative because this is what is required to induce rotation in the $-\that$ direction, which corresponds to a positive electric field (the Brillouin limit arises only with outward electric field).
The second species has the same evolution equation, with index 2 instead of 1 and the forcing $F$ removed.

\subsection{Electron dynamics}
Electrons have a similar evolution equation, except for the signs and the forcing term. The electron momentum equations are:
\begin{align}
  &\dot{\gamma}_e = \Omega_e^2 - \gamma_e^2 - \omega_{ce}\parent{\Omega_E+\Omega_e} - \sum_{s\ne e}\nu_{es}\parent{\gamma_e - \gamma_s} - \nu_{e0}\gamma_e\\
  &\dot{\Omega}_e = -\parent{2\Omega_e-\omega_{ce}}\gamma_e  - \sum_{s\ne e}\nu_{es}\parent{\Omega_e - \Omega_s} - \nu_{e0}\Omega_e.
\end{align}
At the lowest order in $m_e/m_p$, we find the electron velocity to be given by
\begin{align}
  \label{eq:Omegae}
  \Omega_e &= -\Omega_E\\
  \label{eq:gammae}
  \gamma_e &= \sum_{s\ne e}\frac{\nu_{es}}{\omega_{ce}}\parent{\Omega_e - \Omega_s} +\frac{\nu_{e0}}{\omega_{ce}}\Omega_e.
\end{align}
Recall that in general $\nu_{ei}/\nu_{ie}\sim m_p/m_e$, so that $\nu_{es}/\omega_{ce}$ can be considered of order 0 in the electron to ion mass ratio. Equation \refeq{Omegae} indicates that for electrons, the small inertia means the centrifugal and Coriolis forces play a minor role, and the electrons simply rotate at the electric drift frequency. Equation \refeq{gammae} expresses the fact that when the electron forces are balanced, the azimuthal magnetic force associated with the radial electron flux is balanced by the friction with the other species.

\subsection{\label{subsec:efield}Electric field and density dynamics}

To close the system, we need the equation of evolution of the electric field, $\epsilon_0\dive\E = \rho$, where $\rho = n_ee(n_1Z_1/n_e+n_2Z_2/n_e-1)$. Note that on the left hand side, $\epsilon_0$ shows up, rather than $\epsilon_0\epsilon_\perp$. Indeed, the currents associated with the radial velocities $\gamma_1$, $\gamma_2$, $\gamma_e$, are the total currents, taking into account the polarization currents. To see this, it is sufficient to take the cross product of equation \refeq{ionmomeq} with $\B$. At lowest order (neglecting friction) the velocity is $\V_1 = \E\times\B/B_0^2+\F_\mathrm{wave}\times\B/(Z_1n_1eB_0^2)$, and at the next order, the polarization velocity $A_1m_p/(Z_1eB_0^2)\partial_t\E$ appears. We define a dimensionless parameter akin to the electric susceptibility, $\chi_\perp\equiv n_em_p/(\epsilon_0B_0^2)$, and parameterize the ion densities with $\zeta\equiv n_1Z_1/n_e$ and $\eta\equiv n_2Z_2/n_e$. At $t=0$, before the forcing is switched on, we will assume the electric field vanishes, and hence, $\zeta+\eta - 1 = 0$. The electric field drift frequency is, therefore, given by
\begin{align}
  \label{eq:efield}
  \frac{\Omega_E}{\omnorm} = \frac{\chi_\perp}{2}\parent{\zeta+\eta-1}
\end{align}

Numerically, $\chi_\perp$, given by
\begin{align}
  \label{eq:chiperp}
  \chi_\perp = \sci{1.9}{4}\frac{n_e\,\abrack{10^{20}\,\mathrm{m}^{-3}}}{B_0^2\,\abrack{1\,\mathrm{T^2}}},
\end{align}
is large in the plasmas we are considering. However, the following results are very insensitive to its precise value, as long as it is large. When we simulate the dynamical system, the rate of change of momenta is set by the values of $F$ and the friction, but this dynamics is modulated by oscillations at a frequency roughly proportional to $\chi_\perp^{1/2}$. Therefore, it can become expensive numerically, even though we are dealing with ODEs. We will take $\chi_\perp=100$ in the following, which ensures that most of the wave angular momentum is transferred to the plasma motion, with only a small fraction absorbed by the DC field (see appendix~\ref{ap:justif}).

To understand the electric field dynamics, we only need to write the density dynamics. Since we assume all the species to have homogeneous densities initially, the continuity equations for any species is simply $\partial_tn_s = -n_s\dive\V_s=-2n_s\gamma_s$. This yields the evolution equations of $\eta$ and $\zeta$:
\begin{align}
  \label{eq:zetadot}
  \dot{\zeta} &= -2\parent{\gamma_1 - \gamma_e}\zeta \\
  \label{eq:etadot}
  \dot{\eta} &= -2\parent{\gamma_2 - \gamma_e}\eta 
\end{align}

It may be expedient to have an evolution equation of $\Omega_E$, so that one can easily identify the conditions for stationary field. Obviously, the stationarity condition for $\E$ should be that the total radial current vanishes. Indeed, by taking the time derivative of equation \refeq{efield} and using equations \refeq{zetadot}-\refeq{etadot} (and taking into account the variation of $\chi_\perp$), one finds
\begin{eqnarray}
  \label{eq:efielddot}
  \dot{\Omega}_E = -\chi_\perp\omnorm\parent{\gamma_1\zeta + \gamma_2\eta - \gamma_e},
\end{eqnarray}
where the terms in brackets represents the current.

As a final remark regarding the density dynamics, our description hides the fact that an equilibrium solution such that $\gamma_e\ne 0$ or $\gamma_s\ne 0$ necessarily has a varying density, and in that sense is not an equilibrium. If for instance we have $\gamma_2>0$, and the second species is the one we wish to expel, there is no particular conceptual problem, unless that species is the dominant one. 
If the equilibrium is such that for example, $\gamma_e=\gamma_1<0$, a situation that will show up in section~\ref{sec:CollisionalBrillouin}, this means that there is an exponentially increasing (with rate $2|\gamma_e|$) global plasma density. We can still regard the solution as an equilibrium if the final radial velocities have $\gamma\tau_\mathrm{eq}\ll 1$, where $\tau_\mathrm{eq}$ is the time scale of establishment of the equilibrium. In any case, this should not be a big source of worry, as in practice there are sinks and sources of particles, which we have not explicitly taken into account. We are completely neglecting the motion of the plasma in the $\zhat$ direction, but in a real situation where large flows of the order of $\Gamma = 1$ g.s$^{-1}$ are expected, the plasma flows at high velocity along the cylinder, with a plasma source at the entrance of the cylinder and a sink at the exit, and if necessary to collect the unconfined species, on the walls of the cylinder. These questions are beyond the scope of the present paper, but we shall not ignore them completely, because they restrict the validity of some of the conclusions of the present study.

\subsection{Collision frequencies}

To be complete, there remains only to describe our choices for the collision frequencies. The collisions play an important role in this problem, since the electric field drives rotation at the electric drift frequency, but friction with neutrals and between the charged species will damp this motion. A radial current appears, that leads to electric field depletion. Here we only desire to get the correct scalings of mass and charge, without setting the highest importance to the exact value of the friction. We neglect all other transport mechanisms. The easiest choice is to use the friction frequencies for Maxwellians with velocity shifts small compared to the thermal velocity. The friction force that a Maxwellian population of species 1 undergoes when streaming past a Maxwellian population 2 with velocity slip $(\V_1-\V_2)$ is given by
\begin{align}
  \label{eq:friction}
  \RR_{12} &= -\frac{n_1n_2Z_1^2Z_2^2e^4\Lambda}{3\parent{2\pi}^{3/2}\epsilon_0^2T^{3/2}} \sqrt{\frac{m_1m_2}{m_1+m_2}}\parent{\V_1-\V_2} \nonumber\\
  &= -m_1n_1\nu_{12}\parent{\V_1-\V_2},
\end{align}
which defines the collision frequencies $\nu_{12}$ and $\nu_{21}$ for arbitrary mass ratio. 

For the collisions with neutrals, we use the frequency~\cite{Huba2013}
$\nu_{s0} = n_0\sigma_0\sqrt{\frac{2T_s}{m_s}}$, where $n_0$ is the density of neutrals, and $\sigma_0$ is a cross section, taken to be $\sigma_0=\sci{5}{-19}$m$^{2}$.

We assume the plasma to be isothermal at temperature $T$ and define the basic frequencies
\begin{align}
  \label{eq:nu}
  \nu &= \frac{n_ee^4\Lambda}{3(2\pi)^{3/2}\epsilon_0^2m_p^{1/2}T^{3/2}}\sqrt{\frac{m_e}{m_p}}\\
        \label{eq:nun}
      \nu_0 &= n_0\sigma_0\sqrt{\frac{2T}{m_p}}
\end{align}
In our numerical studies, $\nu$ and $\nu_0$ will be adjusted independently, which corresponds to adjusting the temperature and the neutral density. This means that the actual value of $\sigma_0$ does not matter, since any difference (even large) with respect to the value reported in ref.~\cite{Huba2013} can be absorbed in $n_0$. 

We can now define our friction frequencies for our system:
\begin{align}
  \label{eq:nu12}
  &\nu_{12} = \eta\frac{ Z_1^2Z_2}{\sqrt{(m_e/m_p)A_1\parent{1+A_1/A_2}}}\nu\\
  \label{eq:nu21}
  &\nu_{21} = \zeta\frac{ Z_2^2Z_1}{\sqrt{(m_e/m_p)A_2\parent{1+A_2/A_1}}}\nu\\
  \label{eq:nues}
  &\frac{m_e}{m_p}\nu_{e1} = \zeta Z_1 \nu,\quad \frac{m_e}{m_p}\nu_{e2} = \eta Z_2 \nu\\
  \label{eq:nuse}
  &\nu_{se} = \frac{Z_s^2}{A_s}\nu\\
  \label{eq:nu1s}
  &\nu_{s0} = \frac{1}{\sqrt{A_s}}\nu_0,\quad \nu_{e0} = \sqrt{\frac{m_p}{m_e}}\nu_0
\end{align}
where the $s$ subscript designates any of the two ion species and the smallness of $m_e/m_p$ was used. Note that the electron ion friction is in accordance with kinetic theory in a magnetized plasma, which finds exactly equation \refeq{nues} in the perpendicular direction, contrary to the direction parallel to $\B$, where it is reduced by a factor 0.51.

The system of 7 equations, \refeq{gammadot}, \refeq{Omegadot} for both species (without the forcing $F$ for the second species), \refeq{efield}, \refeq{zetadot} and \refeq{etadot}, with $\Omega_e$ and $\gamma_e$ given by equations \refeq{Omegae} and \refeq{gammae} and the collision frequencies by \refeq{nu12}-\refeq{nu1s}, constitutes the closed system of nonlinear ODEs that we shall study in the following sections.

Note that when there is, say, an outflux of electrons, $\gamma_e>0$, the electron density decreases, so that in principle, we should adjust the collision frequencies. This problem leads to a breakdown of our model in the case where the friction between charged species, $\nu$, becomes large (see section~\ref{subsec:ColBrilCharged}). We shall ignore this problem as this is beyond the scope of this paper and barely limits the validity of our conclusions. Indeed, in most cases, the electron radial velocities will remain small (but not negligible because they contribute to the establishment of the electric field).

Finally, note that one can transform the above equations into their normalized versions, suitable for numerical analysis, by merely setting $\omnorm=1$ everywhere. Occasionally in the text, and in all numerical applications, we shall use dimensionless quantities, denoted by bars, for instance $\overline{\nu}=\nu/\omnorm$, $\overline{\nu}_0=\nu_0/\omega_{cp}$.

\section{\label{sec:CollisionalBrillouin}The Brillouin limit in a forced collisional one-species plasma}

In this section, we consider only one species ($\eta=0$) with forcing $F$. We will study the equilibrium of the dynamical system depending on the strength of the forcing.

\subsection{\label{subsec:ColBrilNeutral}Friction with neutrals only}
First, we investigate the case where we retain only the friction with neutrals ($\nu=0$).
The dynamical system becomes
\begin{align}
  \label{eq:dyn1spec1}
  &\dot{\gamma}_1 = \Omega_1^2-\gamma_1^2 + \omega_{c1}\parent{\Omega_1+\Omega_E} - \nu_{10}\gamma_1\\
  \label{eq:dyn1spec2}
  &\dot{\Omega}_1 = -\parent{2\Omega_1+\omega_{c1}}\gamma_1 + \omnorm^2F -\nu_{10}\Omega_1\\
  \label{eq:dyn1spec3}
  &\dot{\zeta} = -2\parent{\gamma_1-\gamma_e}\zeta,
\end{align}
with, from equations \refeq{Omegae}, \refeq{gammae} and \refeq{efield}:
\begin{align}
  \label{eq:nunOmegaE}
  &\Omega_E = \frac{\chi_\perp}{2}\omnorm\parent{\zeta-1}\\
  \label{eq:nungammae}
  &\gamma_e = -\frac{\nu_{e0}}{\omega_{ce}}\Omega_E = -\sqrt{\frac{m_e}{m_p}}\frac{\nu_0}{\omega_{cp}}\Omega_E
\end{align}

We assume the forcing $F$ to be of the order of $\nu_{10}/\omnorm$, which results in $\Omega_1$ being of order $\omnorm$. With $\zeta =\mathcal{O}(1)$, the density equation shows that at equilibrium $\gamma_1\sim (m_e/m_p)\nu_{e0}(\Omega_E/\omnorm)\ll \nu_{10}$ since $\Omega_E/\omnorm=\mathcal{O}(1)$. So we can treat $\gamma_1$ as a small quantity, which means from equation \refeq{dyn1spec2} that at equilibrium
\begin{align}
  \label{eq:frictionOm1}
  \frac{\Omega_1}{\omnorm} \simeq \frac{F}{\nu_{10}/\omnorm}.
\end{align}

\begin{figure}[h!]
  \centering
  \includegraphics[scale=0.4]{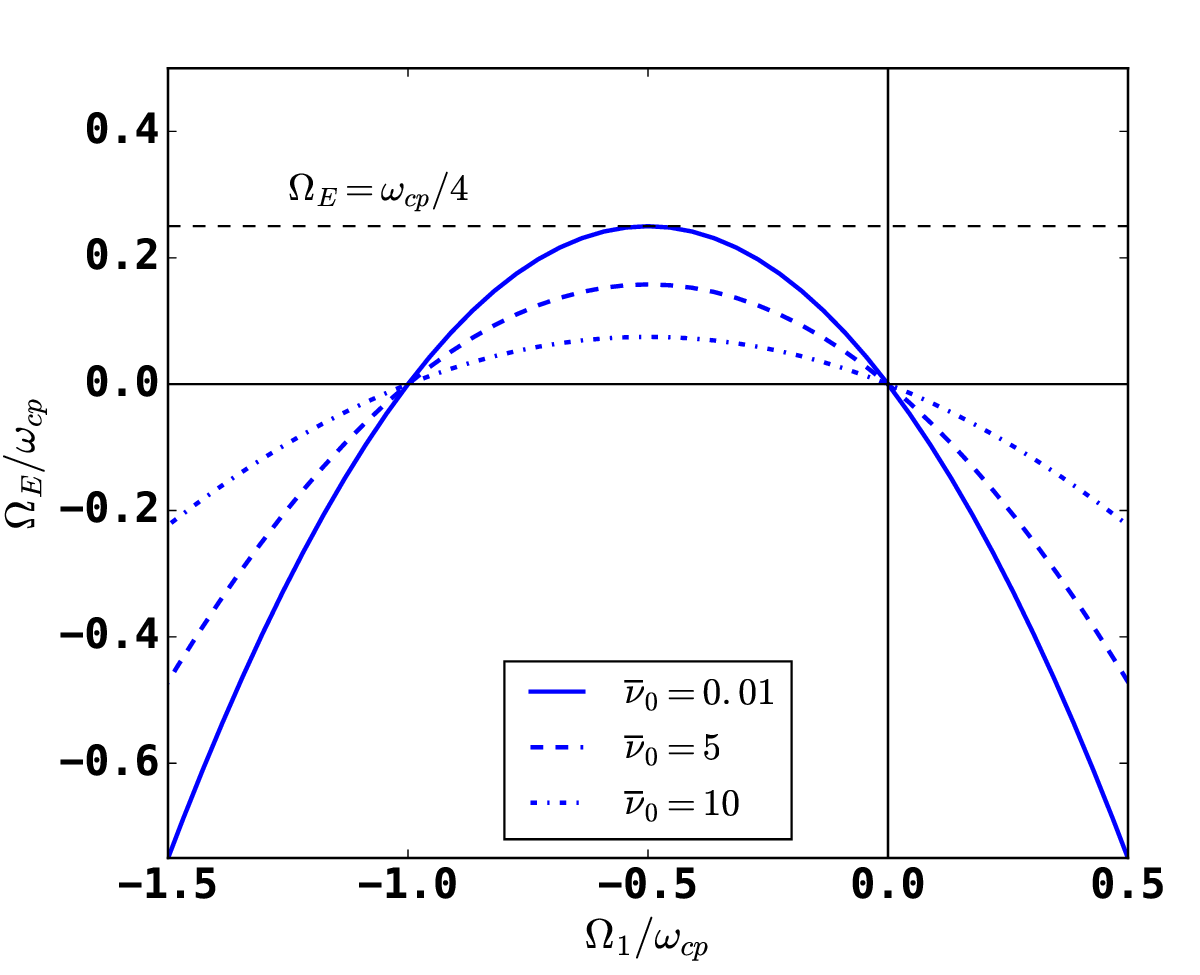}
  \caption{$\Omega_E$ as a function of $\Omega_1$ in the case $A_1=Z_1=1$, for $\overline{\nu}_0=10^{-2}$ (blue solid line), $\overline{\nu}_0=5$ (blue dashed line) and $\overline{\nu}_0 = 10$ (blue dash-dotted line). The black thin horizontal dotted line represents the classical Brillouin limit.}
  \label{fig:Fnu}
\end{figure}
 The meaning of this equation could not be more clear. Since there is injection of momentum at a rate $F$, but also dissipation on neutrals at rest, the equilibrium is simply a balance between the two. More momentum injection means proportionally more rotation. Importantly enough, this is independent of the concept of Brillouin limit. It is important to recall that the Brillouin limit is not a limit in rotation of the ions, but in the electric field that the plasma can sustain! Now the electric field is given by equation \refeq{dyn1spec1}, where we neglect $\gamma_1$:
\begin{align}
  \label{eq:nunBrillouin}
  \Omega_1^2+\omega_{c1}\parent{\Omega_1 + \Omega_E} = 0
\end{align}
This equation is no different than equation \refeq{Brillouinrad} with $\upsilon_r=0$, and the solution is the same. The meaning of the Brillouin limit becomes perfectly clear. As said in the introduction, the electric field establishes as a result of the radial flow caused by the Lorentz force associated with poloidal rotation. The latter arises because of the poloidal forcing. When the forcing is increased, the electric field also increases, until it reaches the Brillouin limit. Increasing the forcing further raises the rotation velocity of the ions, which enter the fast mode regime. If the forcing increases even more, the electric field vanishes and then changes sign, while the ion rotation exceeds the cyclotron frequency. We have to warn, however, that several physical phenomena, \emph{e.g.} possible modification of the wave-plasma coupling, may prevent entering this fast rotation regime.

\begin{figure}[h!]
  \centering
  \includegraphics[scale=0.4]{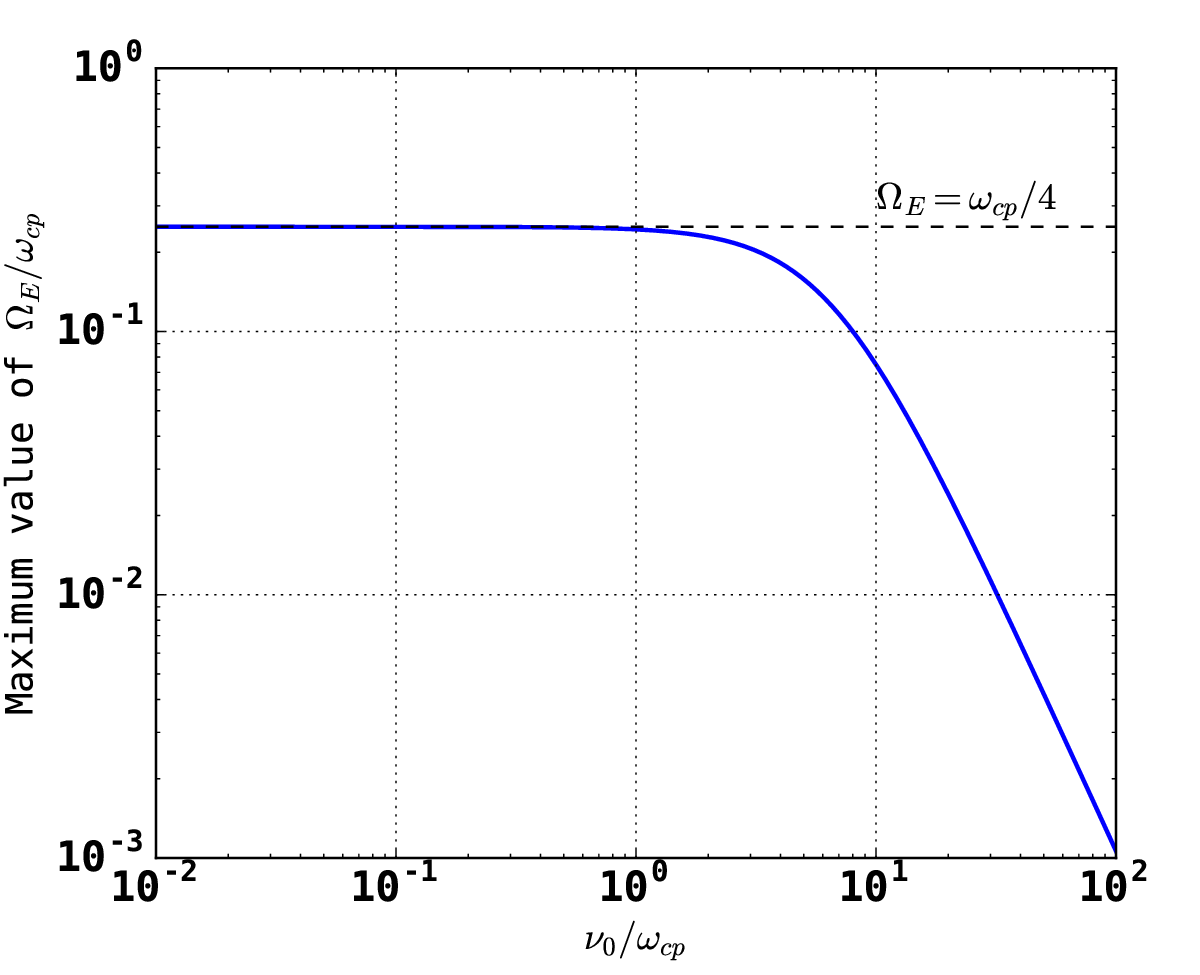}
  \caption{Dependence of the Brillouin limit on the collision frequency with the neutrals $\nu_0$, for $A_1=Z_1=1$}
  \label{fig:omEnu}
\end{figure}

If we do not assume $\nu_0\ll\omnorm$, we can use equations \refeq{dyn1spec1} and \refeq{dyn1spec3} to derive the relation between $\Omega_E$ and $\Omega_1$:
\begin{align} \frac{m_e}{m_p}\overline{\nu}_0^2\Omega_E^2-\parent{\omega_{c1}+\sqrt{\frac{m_e}{m_pA_1}}\overline{\nu}_0^2\omnorm}\Omega_E - \parent{\Omega_1^2+\omega_{c1}\Omega_1}=0,
\end{align}
where $\overline{\nu}_0\equiv\nu_0/\omnorm$. The physically relevant solution for $\Omega_E$ has the minus sign in front of the square root (the other solution has the unphysical property that $\lim_{\nu_0\to0}\Omega_E=\infty$):
\begin{align}
  \label{eq:frictionsol}
  \frac{\Omega_E}{\omega_{c1}} = \frac{1+\sqrt{\frac{m_eA_1}{m_pZ_1^2}}\overline{\nu}_0^2-\sqrt{\parent{1+\sqrt{\frac{m_eA_1}{m_pZ_1^2}}\overline{\nu}_0^2}^2 + 4\frac{m_eA_1^2}{m_pZ_1^2}\overline{\nu}_0^2\parent{\overline{\Omega}_1^2+\frac{Z_1}{A_1}\overline{\Omega}_1}}}{2(m_e/m_p)\overline{\nu}_0^2},
\end{align}
where $\overline{\Omega}_1\equiv\Omega_1/\omnorm$. This expression reduces to \refeq{nunBrillouin} for $\nu_0\ll\omnorm$. The relation~\refeq{frictionsol} is plotted in figure~\ref{fig:Fnu} for $\overline{\nu}_0=0.01$, $\overline{\nu}_0=5$ and $\overline{\nu}_0=10$, with $A_1=Z_1=1$. The position of the point along the curves of figure~\ref{fig:Fnu} is formally set by the solution of equation \refeq{dyn1spec2}. Independently of the value of $\nu_0$, the maximum of $\Omega_E$ is reached for $\Omega_1 = -\omega_{c1}/2$. The resulting dependence between the maximum of $\Omega_E$ and $\nu_0$ is plotted in figure~\ref{fig:omEnu}.

Not only do we not observe a breakdown of the Brillouin limit because of collisions, but in the strongly collisional (or weakly magnetized) regime where $\nu_0$ becomes larger than the ion cyclotron frequency, the maximum value of $\Omega_E$ that can be reached by forcing the plasma rotation decreases as $\nu_0^{-2}$.

\subsection{\label{subsec:ColBrilCharged}Coulomb collisions}

 Now let us consider the case where there are no neutrals, but we take into account the Coulomb collisions between the charged species. With only one species, we only have the electron ion collisions. Using $\Omega_e = -\Omega_E$ and the expressions of the collision frequencies as a function of $\nu$, the dynamical system becomes
\begin{align}
  \label{eq:cdyn1spec1}
  &\dot{\gamma}_1 = \Omega_1^2-\gamma_1^2 + \omega_{c1}\parent{\Omega_1+\Omega_E} - \frac{Z_1^2}{A_1}\nu\parent{\gamma_1-\gamma_e}\\
  \label{eq:cdyn1spec2}
  &\dot{\Omega}_1 = -\parent{2\Omega_1+\omega_{c1}}\gamma_1 + \omnorm^2F -\frac{Z_1^2}{A_1}\nu\parent{\Omega_1+\Omega_E}\\
  \label{eq:cdyn1spec3}
  &\dot{\zeta} = -2\parent{\gamma_1-\gamma_e}\zeta,
\end{align}
with, from equations \refeq{Omegae}, \refeq{gammae} and \refeq{efield}:
\begin{align}
  \label{eq:nuOmegaE}
  &\Omega_E = \frac{\chi_\perp}{2}\omnorm\parent{\zeta-1}\\
  \label{eq:nugammae}
  &\gamma_e = \frac{\nu_{e1}}{\omega_{ce}}\parent{\Omega_e - \Omega_1} = -\zeta Z_1\frac{\nu}{\omnorm}\parent{\Omega_E+\Omega_1}
\end{align}
Some conclusions from the previous subsection still hold, but there are some differences, because contrary to neutrals, electrons are also set into motion by the electric field, so that the slip between the two frictional species is smaller. For small friction, the previous argument about the smallness of $\gamma_1$ remains valid, and we recover equation~\refeq{nunBrillouin}. The relation between $\Omega_1$ and $F$ is, however, strongly modified. Using the notation $\Xi\equiv \Omega_1+\Omega_E$, equation~\refeq{cdyn1spec1} where we neglect $\gamma_1^2$ reads
\begin{align}
  &\Xi = -\frac{\Omega_1^2}{\omega_{c1}}.
\end{align}
Note that the radial friction term disappears because at equilibrium $\gamma_1=\gamma_e$ according to equation~\refeq{cdyn1spec3}. Also, by replacing $\gamma_1$ with $\gamma_e$ from equation~\refeq{nugammae} in equation~\refeq{cdyn1spec2}, we find after multiplying by $\omega_{cp}^2/(A_1\nu)$:
\begin{align}
  \abrack{\parent{2\Omega_1+\omega_{c1}}\zeta - \omega_{c1}}\omega_{c1}\Xi = \omnorm^3\left|\frac{F}{A_1\nu/\omnorm}\right|.
\end{align}
Using that $\zeta-1 = 2(\Omega_E/\omnorm)/\chi_\perp\ll1$, we find that
\begin{align}
  \label{eq:Fsurnuscaling}
  \frac{\Omega_1}{\omega_{cp}} \simeq -\left|\frac{F}{2A_1\overline{\nu}}\right|^{1/3},
\end{align}
where $\overline{\nu}\equiv\nu/\omnorm$.
Compared to the friction with neutrals, the scaling of $\Omega_1$ with $F/\overline{\nu}$ is smaller, as expected. This result holds only for $\nu\ll\omnorm$, because we have neglected the $\gamma_1^2$ term from equation \refeq{cdyn1spec1} . When $\nu\ll\omnorm$, the only relevant parameter is the ratio $|F/\overline{\nu}|$, so the only result of having larger friction is that one has to proportionally increase the friction in order to reach the same level of rotation and electric field.

We will only briefly consider the case $\nu>\omnorm$. It is not really relevant for the plasmas we are interested in. Indeed
\begin{align}
  \frac{\nu}{\omnorm} \simeq \sci{1.8}{-3} \frac{n_e\abrack{10^{19}\mathrm{m}^{-3}}}{B_0\abrack{1\mathrm{T}}T^{3/2}\abrack{1\mathrm{eV}}},
\end{align}
where we have used the value $\Lambda=11$, assumes rather small values in practice. Retaining $\gamma_1=\gamma_e$ in the radial equation gives $\Omega_1$ as a function of $\Omega_E$:
\begin{align}
  \label{eq:OmElargenu}
  \frac{\Omega_E}{\omnorm} = \frac{1-\sqrt{1 + 4A_1^2\overline{\nu}^2\Omega_1^2/\omnorm^2}}{2Z_1A_1\overline{\nu}^2} - \frac{\Omega_1}{\omnorm},
\end{align}  
where the minus sign in front of the square root comes from the requirement that $\Omega_E$ vanish when $\Omega_1=0$. It can be easily seen that the case $\nu>\omnorm$ makes the whole device unusable. The radial velocity can be obtained in the limit $\nu\to\infty$ by plugging the solution \refeq{OmElargenu} into \refeq{nugammae} and using $\gamma_1=\gamma_e$, which results in
\begin{align}
  \gamma_1 = \zeta|\Omega_1|\simeq |\Omega_1|,
\end{align}
since $\zeta\simeq 1$. This means the plasma is radially expelled in the time scale of one angular rotation. Therefore, it is required, for the Ohkawa filter to work, to keep the Coulomb collision frequency smaller than the ion cyclotron time. There is no such requirement regarding the neutral collision frequency.


\section{\label{sec:BrillouinBreakdown} The interpretation of the Brillouin limit with collisions}

The two previous subsections have shown that the relation between $\Omega_E$ and $\Omega_1$, equation~\refeq{nunBrillouin}, leading to the Brillouin limit of $\Omega_E = \omega_{c1}/4$, remains valid when collisions are taken into account. When it has to be corrected for, in the regime of large collisions with neutrals, it leads to a strong decrease of the Brillouin limit, rather than a disappearance of that limit (recall figure~\ref{fig:omEnu}). How can we reconcile this with ref~\cite{raxBreakdownBrillouinLimit2015}, which proves the breakdown of the Brillouin limit in presence of collisions with neutrals? In section~\ref{subsec:fluid}, we first analyzed a collisionless plasma and found the Brillouin limit. When we introduced the collisions, we insisted that it is necessary to explicitly take into account the forcing term of the plasma, because the electric field is generated indirectly by the radial motion induced by the inertial, Lorentz and friction forces in presence of the azimuthal forcing. If we fail to do so, there is an inconsistency, because we assume there is an electric field at equilibrium, while it should naturally fade away under the action of dissipation. On the contrary, in ref~\cite{raxBreakdownBrillouinLimit2015}, if one assumes the electric field is generated by biased electrodes, one can treat $\Omega_E$ as an external parameter (in fact, the electric field arises from the electron motion in the $z$ direction, toward or coming from the electrodes). Then, one can consider only equations~\refeq{dyn1spec1}-\refeq{dyn1spec2} without \refeq{dyn1spec3} (which gives the electric field dynamics in our case), and without the azimuthal forcing:
\begin{align}
  \label{eq:wBrillouin1}
  &\Omega_1^2-\gamma_1^2 + \omega_{c1}\parent{\Omega_1+\Omega_E} - \nu_{10}\gamma_1=0\\
  \label{eq:wBrillouin2}
  &-\parent{2\Omega_1+\omega_{c1}}\gamma_1 - \nu_{10}\Omega_1=0.
\end{align}
\begin{figure}[t]
  \centering
  \includegraphics[scale=0.4]{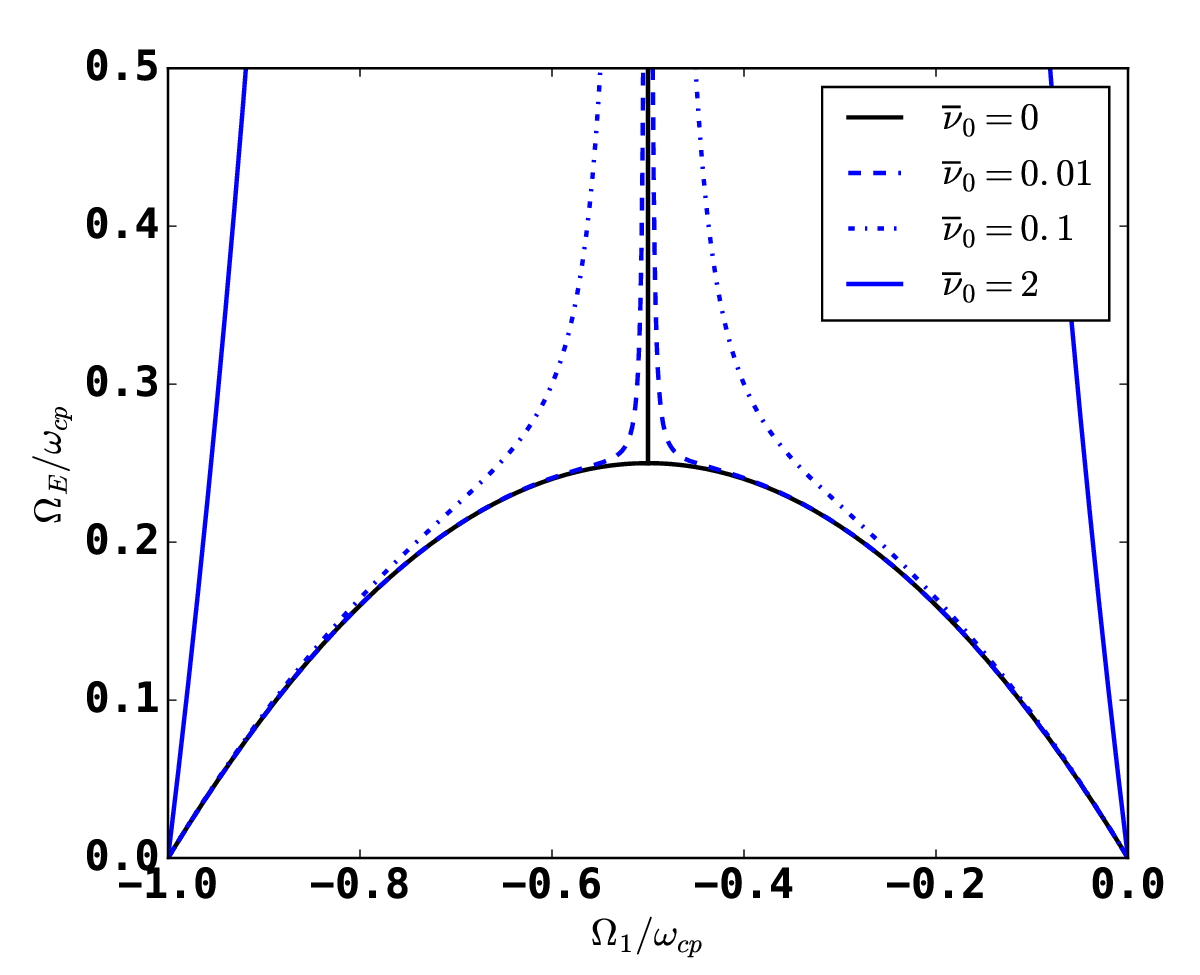}
  \caption{Relation $\Omega_E = f(\Omega_1)$ from equation~\refeq{breakdown} for $Z_1=A_1=1$, with $\overline{\nu}_0=0$ (black solid line), $\overline{\nu}_0=0.01$ (dashed line), $\overline{\nu}_0=0.1$ (dash-dotted line) and $\overline{\nu}_0=2$ (solid line).}
  \label{fig:breakdown}
\end{figure}
\begin{figure}[t]
  \centering
  \includegraphics[scale=0.4]{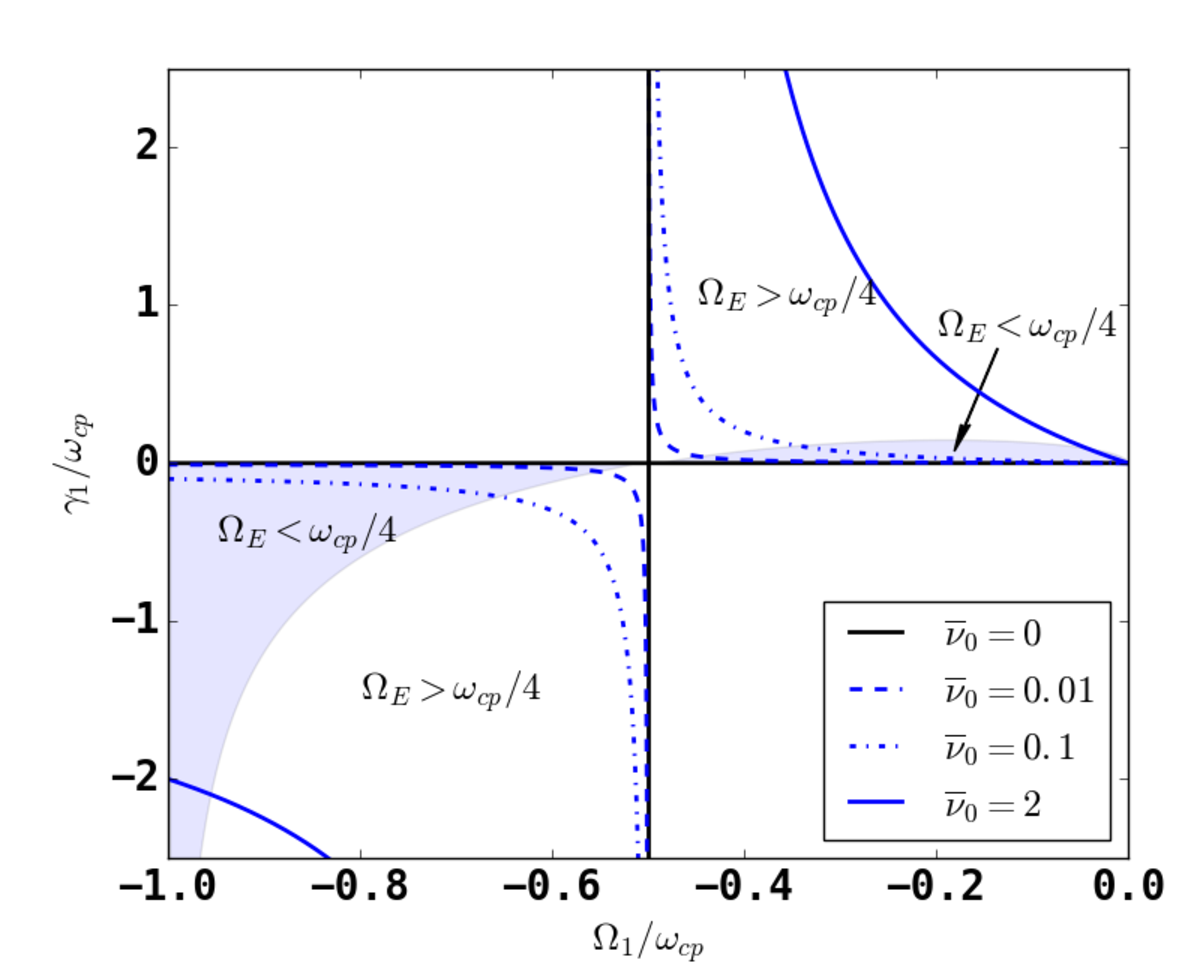}
  \caption{Radial velocity solution of equation~\refeq{breakdown} as a function of $\Omega_1$, for $Z_1=A_1=1$, with $\overline{\nu}_0=0$ (black solid line), $\overline{\nu}_0=0.01$ (dashed line), $\overline{\nu}_0=0.1$ (dash-dotted line), $\overline{\nu}_0=2$ (solid line). The highlighted zone is where the electric field is below the Brillouin limit. }
  \label{fig:breakdownGam}
\end{figure}
Now one is merely interested whether there is a solution for $\Omega_1$ for any value of $\Omega_E$. Instead of equating $\gamma_1$ and $\gamma_e$, as required by the electric field equilibrium, one takes $\gamma_1$ from equation~\refeq{wBrillouin2} and plugs it into equation~\refeq{wBrillouin1}. Reformulating the problem with the variable $\alpha = -(2\Omega_1+\omega_{c1})/\nu_{10}$ (incidentally, this variable has a physical meaning since $2\Omega_1+\omega_{c1}$ is the effective cyclotron frequency in the rotating frame\cite{lehnert1964dynamics}), one obtains a second order equation for $\alpha^2$:
\begin{align}
  \label{eq:breakdown}
  \alpha^4+\alpha^2\abrack{1-\frac{\omega_{c1}^2}{\nu_{10}^2}\parent{1 - 4\frac{\Omega_E}{\omega_{c1}}}} - \frac{\omega_{c1}^2}{\nu_{10}^2}=0
\end{align}
This is precisely equation (12) from Ref.~\cite{gueroultCentrifugalInstabilityRegime2017}, if $Z_1=A_1=1$ and with the substitutions
\begin{align}
  &\frac{\nu_{10}}{\omnorm}\longrightarrow\frac{\nu_i}{\Omega_i}\\
  &\frac{\Omega_E}{\omnorm}\longrightarrow-\frac{\phi'}{rB_0\Omega_i}.
\end{align}
The relation \refeq{breakdown} in the $A_1=Z_1=1$ case is plotted in Fig~\ref{fig:breakdown} for $\overline{\nu}_0=0$, $\overline{\nu}_0=0.01$, $\overline{\nu}_0=0.1$ and $\overline{\nu}_0=2$. It is this relation that is interpreted as a breakdown of the Brillouin limit (in the case where the electric field is generated by biased electrodes), because it has a solution $\alpha^2>0$ for any value of $\Omega_E$. In our case however, this would be inconsistent if there is only one species in the plasma, because of the azimuthal forcing. In our case, figure~\ref{fig:breakdown} ought to be replaced with figure~\ref{fig:Fnu}. We want to emphasize, however, that the solutions of \refeq{breakdown} are perfectly acceptable if the considered ion species plays no part in the establishment of the electric field. In other words, if we consider a trace impurity. 

There is, however, another important consequence of equation~\refeq{breakdown}. Each solution for $\alpha$, which gives $\Omega_1$, also comes with a radial velocity $\gamma_1 = \Omega_1/\alpha$. The solutions are plotted in figure~\ref{fig:breakdownGam}. We have highlighted the zone where $\Omega_E$ is less than the Brillouin limit. It is clearly seen that the sign of the radial velocity depends on whether the solution is in the fast or the slow mode. The slow mode is on the right of the figure, and the radial velocities are positive, even when $\Omega_E<\Ommax{1}$. This means that a trace species can be expelled radially even if the electric field is below that species' Brillouin limit.

\section{\label{sec:trace}Dynamics of a trace impurity with respect to the Brillouin limit}

\subsection{\label{subsec:neutralcol}Collisions with neutrals}

We have already argued in the previous section that if we are considering a trace impurity, the equations \refeq{wBrillouin1} and \refeq{wBrillouin2}, or \refeq{breakdown}, apply (with index 1 replaced by index 2). The electric field is set by the dynamics of the first species, while it becomes sufficient to read on Figs.~\ref{fig:breakdown} and \ref{fig:breakdownGam} the values of $\Omega_2$ and $\gamma_2$. An important comment is in order, though. In the absence of electric field, the impurity naturally starts at rest with $\Omega_2=0$ on the right of the figures, and we see that when $\Omega_E$ is increased (let's say, the forcing of the first species is increased), $\Omega_2$ increases but cannot cross the line $\Omega_2=-\omega_{c2}/2$. In other words, it can never reach the fast mode where $\Omega_2<-\omega_{c2}/2$. Hence, the left part of figures~\ref{fig:breakdown} and \ref{fig:breakdownGam} is basically irrelevant. An important consequence is that when we consider only the collisions with the neutrals, $\gamma_2$ is always positive (ejection), \emph{whether it is a heavy impurity or a light impurity}. We will come back to this point.

\subsection{\label{subsec:chargedcol} Coulomb collisions}

We will now study the dynamics of a trace impurity when Coulomb collisions are the only source of friction. The results are different whether we consider an impurity that is heavier or lighter than the first (forced) species.

\subsubsection{Heavy impurity}
\begin{figure}[h!]
  \centering
  \includegraphics[scale=0.4]{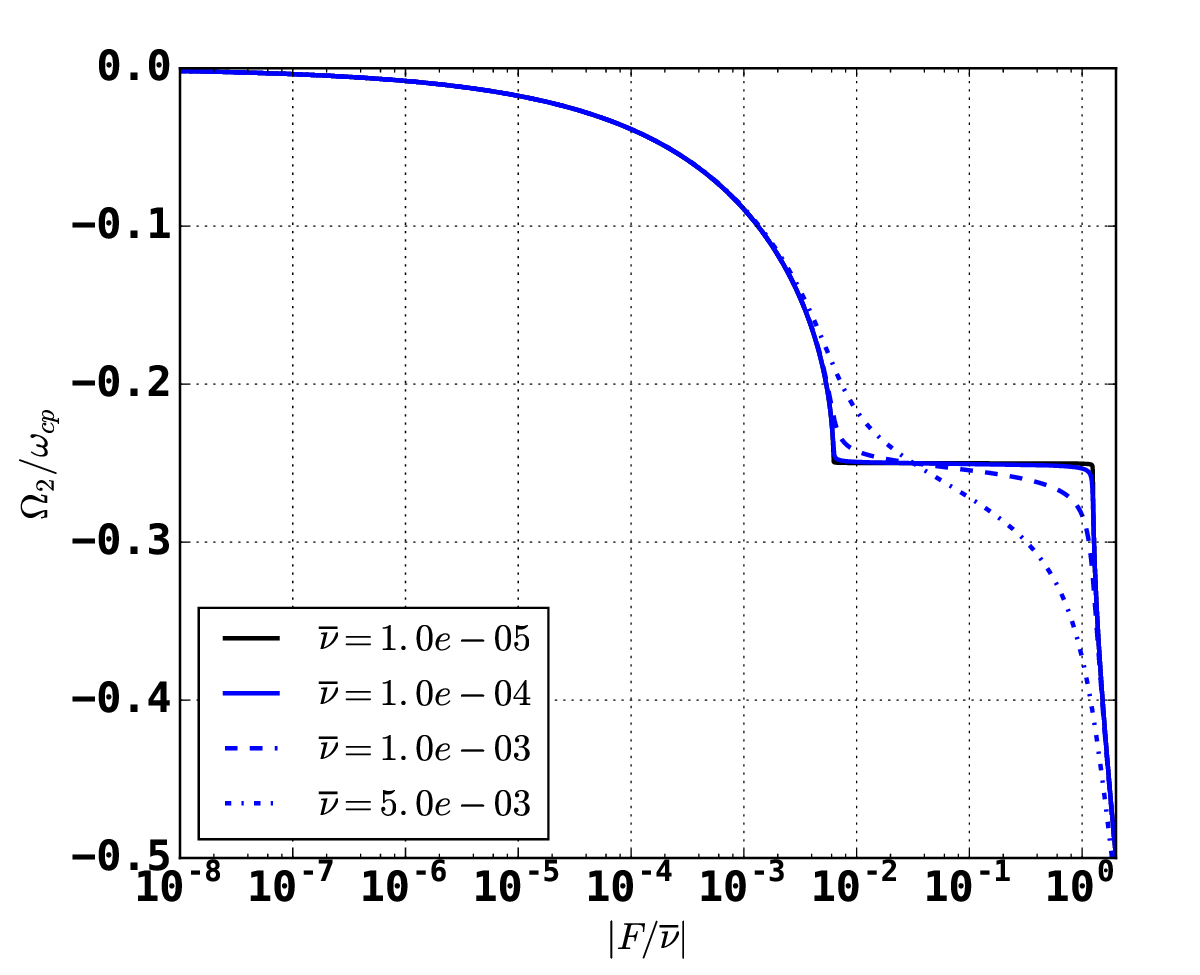}
  \caption{Relation between $|F/\overline{\nu}|$ and $\Omega_2$ when $A_1=Z_1=Z_2=1$ and $A_2=2$, for $\overline{\nu}=10^{-5}$ (black solid line), $\overline{\nu}=10^{-4}$ (solid blue line), $\overline{\nu}=10^{-3}$ (dashed blue line), and $\overline{\nu}=\sci{5}{-3}$ (dash-dotted blue line). Recall that $F<0$.}
  \label{fig:breakdown_charged_For}
\end{figure}
\begin{figure}[h!]
  \centering
  \includegraphics[scale=0.4]{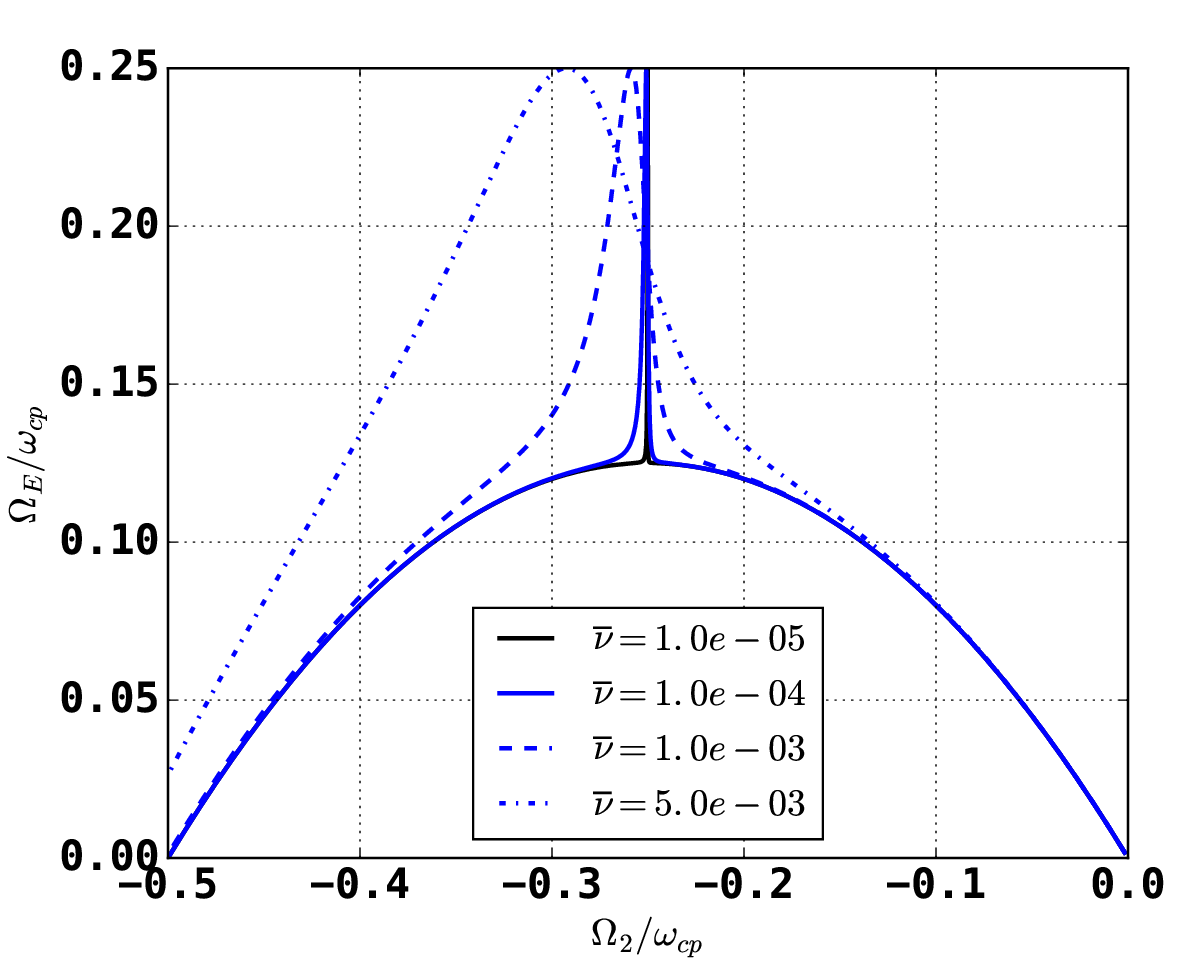}
  \caption{Relation between $\Omega_2$ and $\Omega_E$ when $A_1=Z_1=Z_2=1$ and $A_2=2$, for $\overline{\nu}=10^{-5}$ (black solid line), $\overline{\nu}=10^{-4}$ (solid blue line), $\overline{\nu}=10^{-3}$ (dashed blue line), and $\overline{\nu}=\sci{5}{-3}$ (dash-dotted blue line). }
  \label{fig:breakdown_charged_Om}
\end{figure}
\begin{figure}[h!]
  \centering
  \includegraphics[scale=0.4]{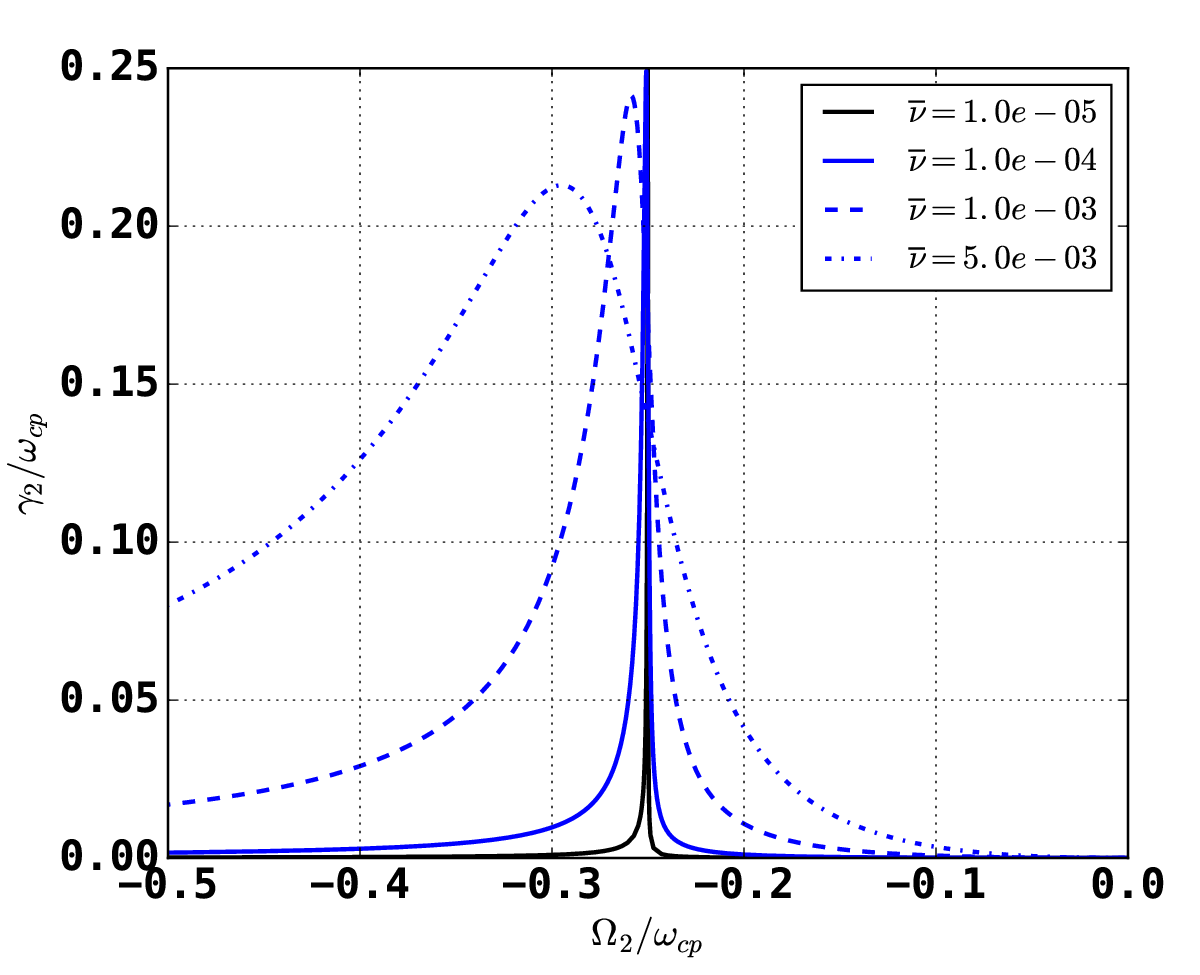}
  \caption{Relation between $\Omega_2$ and $\gamma_2$ when $A_1=Z_1=Z_2=1$ and $A_2=2$, for $\overline{\nu}=10^{-5}$ (black solid line), $\overline{\nu}=10^{-4}$ (solid blue line), $\overline{\nu}=10^{-3}$ (dashed blue line), and $\overline{\nu}=\sci{5}{-3}$ (dash-dotted blue line).}
  \label{fig:breakdown_charged_Gam}
\end{figure}

The equations now involve a coupling between the two species through friction, other than through the sole electric field as was the case for collisions with neutrals. We assume there is a forced ion (numbered 1) with a trace impurity (numbered 2). We will produce figures equivalent to \ref{fig:breakdown} and \ref{fig:breakdownGam} for the second species. We have already seen how to compute $\Omega_1$ and $\Omega_E$ as a function of $F/\nu$ for $\nu\ll \omnorm$ (equation~\refeq{Fsurnuscaling}). Then we can solve for $\Omega_2$ and $\gamma_2$ numerically. Since the second trace species is heavier than the first, it has $\Ommax{2}<\Ommax{1}$, and we will be able to force the system at $\Omega_E>\Ommax{2}$. We simply take $A_1=Z_1=1$ and $A_2=2$, $Z_2=1$, then $\Ommax{1} = 1/4\omnorm$, while $\Ommax{2} = 1/8\omnorm$. 

The results can be seen in figures~\ref{fig:breakdown_charged_For}, \ref{fig:breakdown_charged_Om} and \ref{fig:breakdown_charged_Gam}. There are similarities as well as differences with the case of neutral collisions, figures~\ref{fig:breakdown} and \ref{fig:breakdownGam}. Fig~\ref{fig:breakdown_charged_For} shows that in the limit $\nu\to0$ (but $F/\nu$ remaining finite), there is a wide range of values for the forcing where $\Omega_2$ is locked at $\Omega_2 = -\omega_{c2}/2$. Recall that the forcing determines $\Omega_1$, and $\Omega_E$, through the balance between inertial, Lorentz and electron-ion friction forces. Thus, the domain where $\Omega_2 = -\omega_{c2}/2$ is the domain where the forcing is such that $\Omega_E>\Ommax{2}$. For these values of the forcing, $\gamma_2$ is positive, otherwise it vanishes. When $\nu$ becomes finite, this sharp behaviour is smoothed, as we have seen in the case of collisions with neutrals. But the difference with figures~\ref{fig:breakdown} and \ref{fig:breakdownGam} is that the solutions are asymmetric with respect to $\Omega_2=-\omega_{c2}/2$, and that $\gamma_2$ never becomes negative. Mathematically, there is another solution to the nonlinear system, that is the exact symmetric of the physical one with respect to the axis $\Omega=-\omega_{c2}/2$. More precisely, it is symmetric for $\Omega_2$ and antisymmetric for $\gamma_2$, so that $\gamma_2<0$ for that solution. It is unphysical because it has $\Omega_2$ decreasing when $F$ increases, with $\lim_{F\to0}\Omega_2=-\omega_{c2}/2$.

An important difference, also to be noted, is that now the fast mode of the second species, where $|\Omega_2|>\omega_{c2}/2$, becomes accessible, contrary to the case of neutral collisions (see section~\ref{subsec:neutralcol}).

\subsubsection{Light impurity}
\begin{figure}[h!]
  \centering
  \includegraphics[scale=0.4]{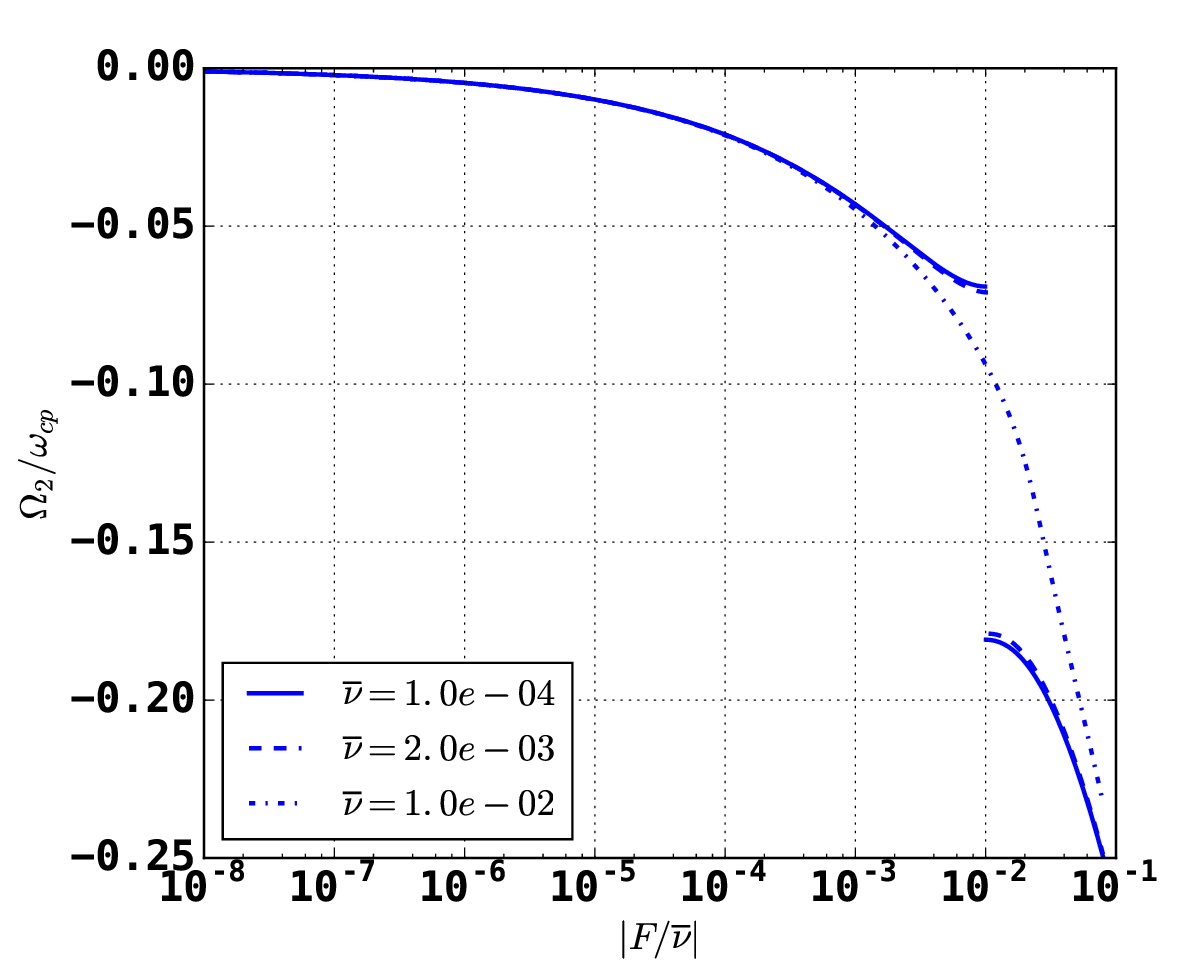}
  \caption{Relation between $|F/\overline{\nu}|$ and $\Omega_2$ when $Z_1=Z_2=1$, $A_1=5$ and $A_2=4$, for $\overline{\nu}=10^{-5}$ (black solid line), $\overline{\nu}=10^{-4}$ (solid blue line), $\overline{\nu}=\sci{2}{-3}$ (dashed blue line), and $\overline{\nu}=\sci{5}{-3}$ (dash-dotted blue line). Recall that $F<0$.}
  \label{fig:breakdown_charged_For_light}
\end{figure}
\begin{figure}[h!]
  \centering
  \includegraphics[scale=0.4]{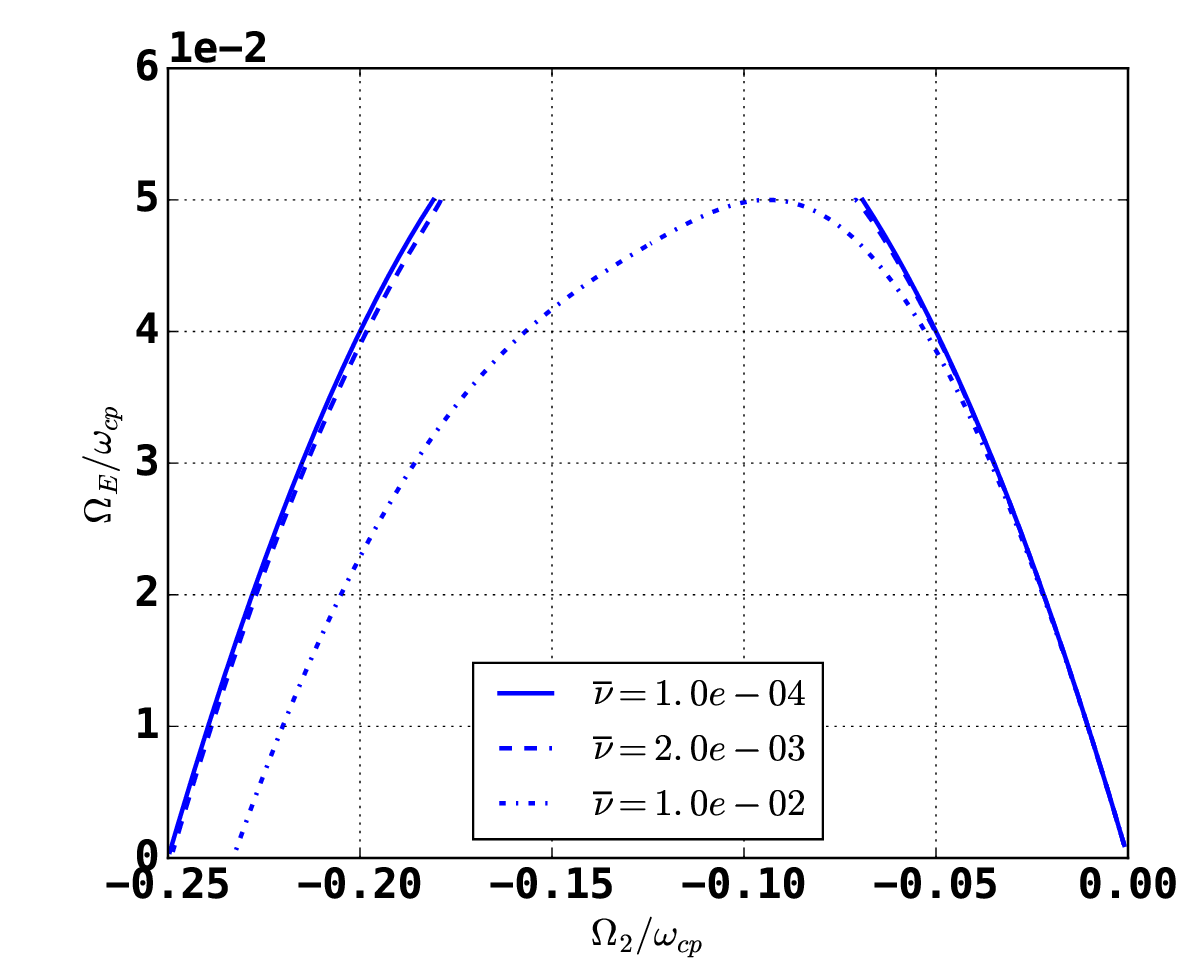}
  \caption{Relation between $\Omega_2$ and $\Omega_E$ when $Z_1=Z_2=1$, $A_1=5$ and $A_2=4$, for $\overline{\nu}=10^{-5}$ (black solid line), $\overline{\nu}=10^{-4}$ (solid blue line), $\overline{\nu}=\sci{2}{-3}$ (dashed blue line), and $\overline{\nu}=\sci{5}{-3}$ (dash-dotted blue line). }
  \label{fig:breakdown_charged_Om_light}
\end{figure}
\begin{figure}[h!]
  \centering
  \includegraphics[scale=0.4]{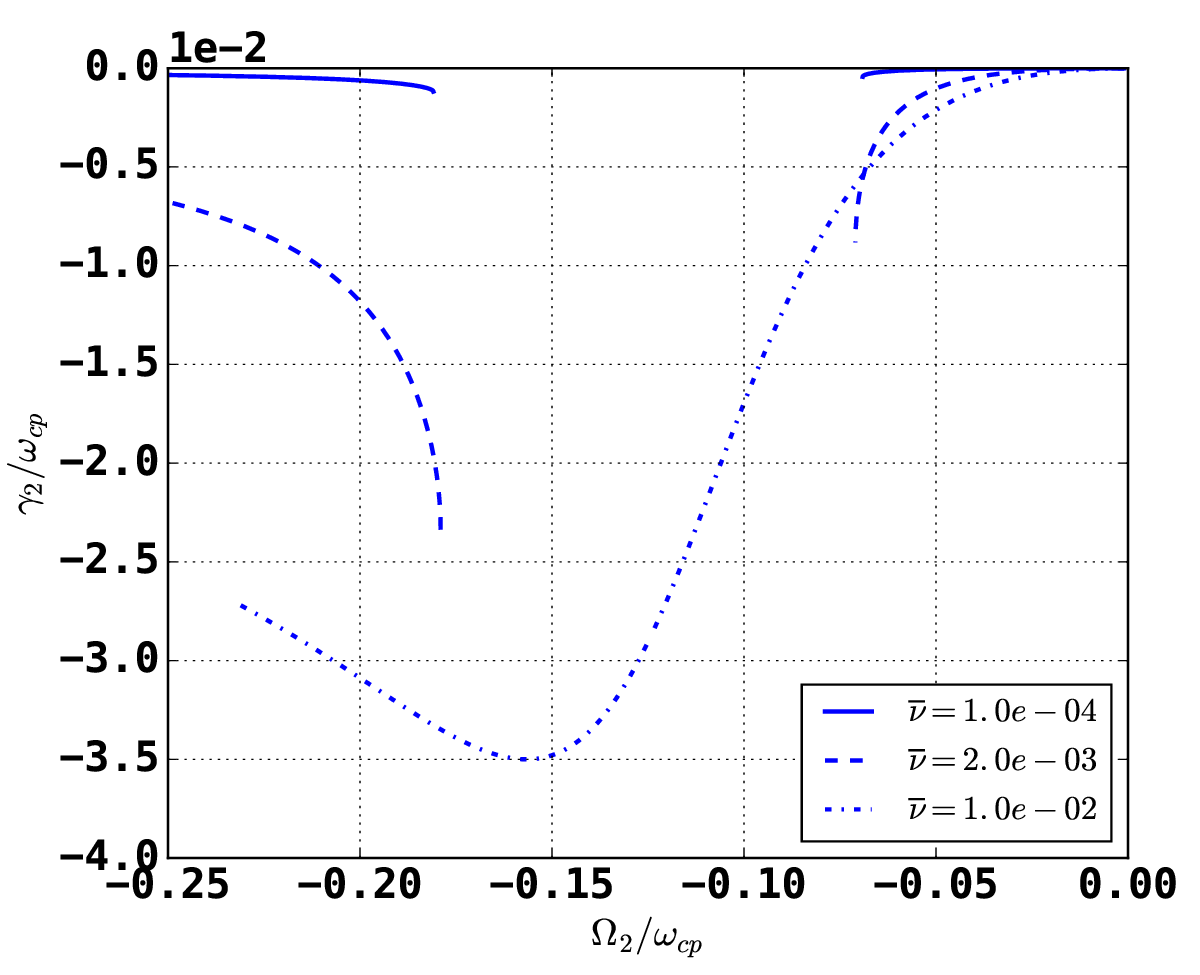}
  \caption{Relation between $\Omega_2$ and $\gamma_2$ when $Z_1=Z_2=1$, $A_1=5$ and $A_2=4$, for $\overline{\nu}=10^{-5}$ (black solid line), $\overline{\nu}=10^{-4}$ (solid blue line), $\overline{\nu}=\sci{2}{-3}$ (dashed blue line), and $\overline{\nu}=\sci{5}{-3}$ (dash-dotted blue line).}
  \label{fig:breakdown_charged_Gam_light}
\end{figure}

We can do the same exercise when the second species is lighter than the first. The results change in significant ways, as can be seen in Figs.~\ref{fig:breakdown_charged_For_light}, \ref{fig:breakdown_charged_Om_light} and \ref{fig:breakdown_charged_Gam_light}, where $A_1=5$, $A_2=4$, and $Z_1=Z_2=1$. The electric field can only be forced up to $\Omega_E = \Ommax{1}<\Ommax{2}$. For small $\nu$, as we increase the forcing, there is a jump in the poloidal frequency $\Omega_2$. This is easily understood because the relation $\Omega_2^2+\omega_{c2}(\Omega_2+\Omega_E)\approx 0$ holds and friction with the first species drives $\Omega_2$ while at the same time $\Omega_E$ cannot surpass (or even reach) $\Ommax{2}$. The usual relation between $\Omega_2$ and $\Omega_E$ explains the parabola branches in figure~\ref{fig:breakdown_charged_Om_light}, and the existence of a jump for a certain value of the forcing is readily understood. Another important difference is that now $\gamma_2$ is always negative.

Note that the results we have just discussed apply strictly speaking only to the case of a trace impurity. Here, the behaviour of the first species is determined by the forcing and the friction with the electrons, while that of the second species is determined by the friction with electrons and with the first ion species.

\subsection{\label{subsec:paramspace}Exploration of the $(\nu,\nu_0)$ parameter space}

So far, we have only considered the case of either only friction with neutrals, or only Coulomb collisions. When both $\nu$ and $\nu_0$ are nonzero, it is all the more necessary to carry out numerical parameter studies because of the intricacies of the nonlinear system. We can first sum up what we have already seen so far. When the trace impurity is heavier than the dominant species, whether due to neutral or Coulombian collisions, it has $\gamma_2>0$. Therefore, we can assume it will always have $\gamma_2>0$ whatever the values of $\nu$ and $\nu_0$. When the trace impurity is lighter however, collisions with the neutrals lead to $\gamma_2>0$, while Coulombian collisions lead to $\gamma_2<0$. Therefore, we expect the presence of a boundary in the $(\nu,\nu_0)$ space between a region dominated by neutral collisions where $\gamma_2>0$ and a region dominated by Coulombian collisions where $\gamma_2<0$.

We draw the reader's attention on another point. We have seen in section \ref{sec:CollisionalBrillouin} that for collisions with neutrals, at equilibrium $\gamma_1=\gamma_e\propto -\Omega_E<0$ whereas in the case of Coulombian collisions, $\gamma_1=\gamma_e\propto -(\Omega_E+\Omega_1)>0$. The question, then, is not only to know if $\gamma_2$ is positive, but aso if $\gamma_2>\gamma_1$.

In both cases of heavy and light trace impurity, our algorithm is as follows. First we fix $A_1$, $Z_1$ and $A_2$, $Z_2$. Then, for each couple of values for $(\nu,\nu_0)$, we vary the forcing $F$. For each value of the forcing we find numerically the equilibrium solution of the system. Since the second species is a trace, the problem is simplified. First we can find $\gamma_1$, $\Omega_1$ and $\Omega_E$. Then, plugging these values in the equations for the second species, we find $\gamma_2$ and $\Omega_2$. The only subtlety is that there are two solutions for $\gamma_2$, $\Omega_2$, and we have to make sure to obtain the physical one, such that $\ud\Omega_2/\ud F >0$. This allows to find the value of $F$ so that the electric field is maximum. We can then plot the values of $\gamma_1$ and $\gamma_2$ for this value of the forcing. Indeed we assume that the most favorable situation, the one where the second species is most likely to have large radial velocities, is when the electric field is maximum, in accordance with the basic physics of the Brillouin limit.

\subsubsection{\label{subsubsec:paramspace_heavy} Heavy impurity}
\begin{figure}[t!]
  \centering
  \includegraphics[scale=0.38]{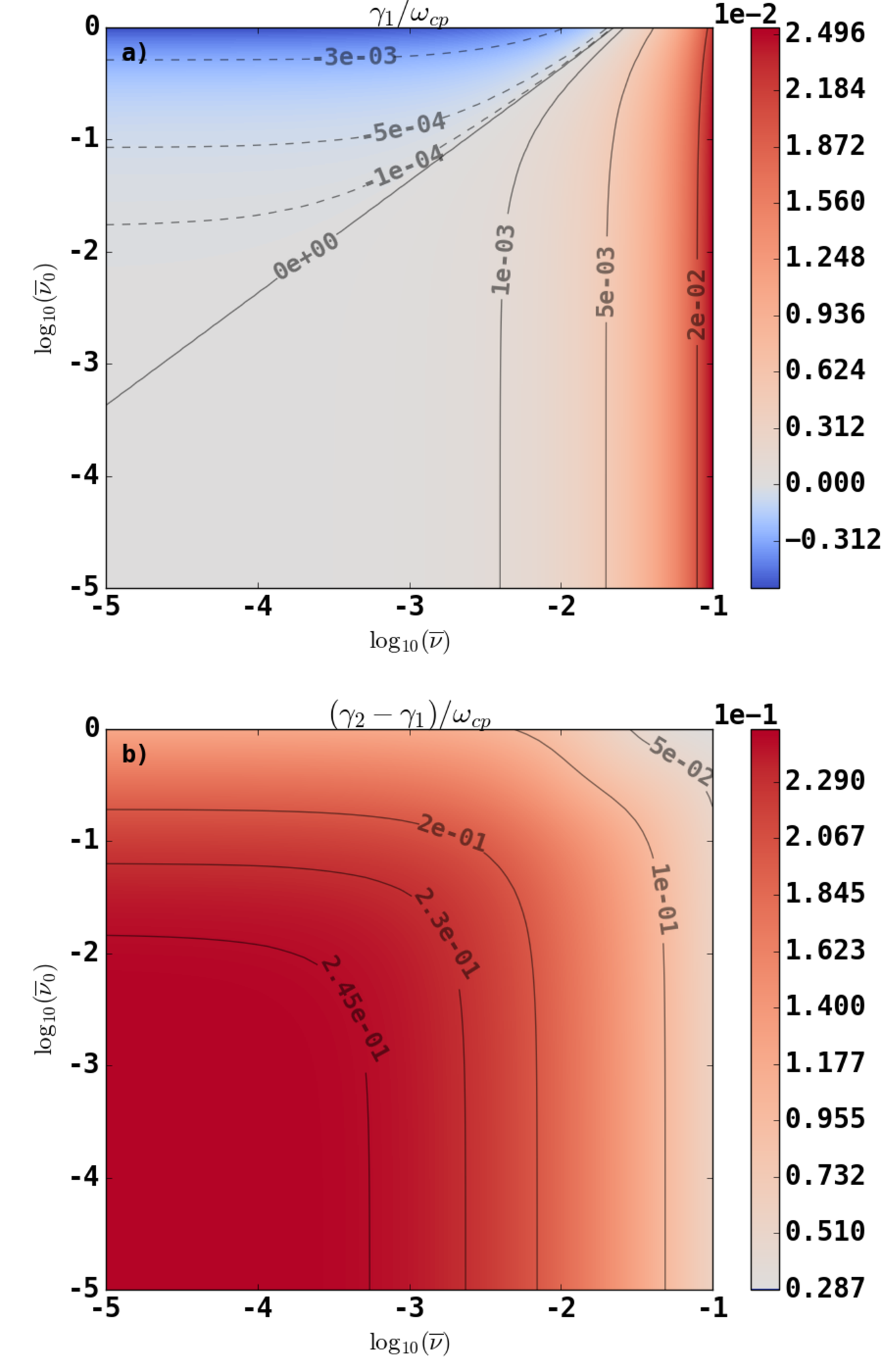}
  \caption{Dependence of $\gamma_1$ (a) and $\gamma_2-\gamma_1$ (b) in the $(\nu,\nu_0)$ parameter space with $Z_1=Z_2=1$, $A_1=1$, $A_2=2$, when the first species is forced so that $\Omega_E=\Ommax{1}$. }
  \label{fig:HD}
\end{figure}
\begin{figure}[t!]
  \centering
  \includegraphics[scale=0.38]{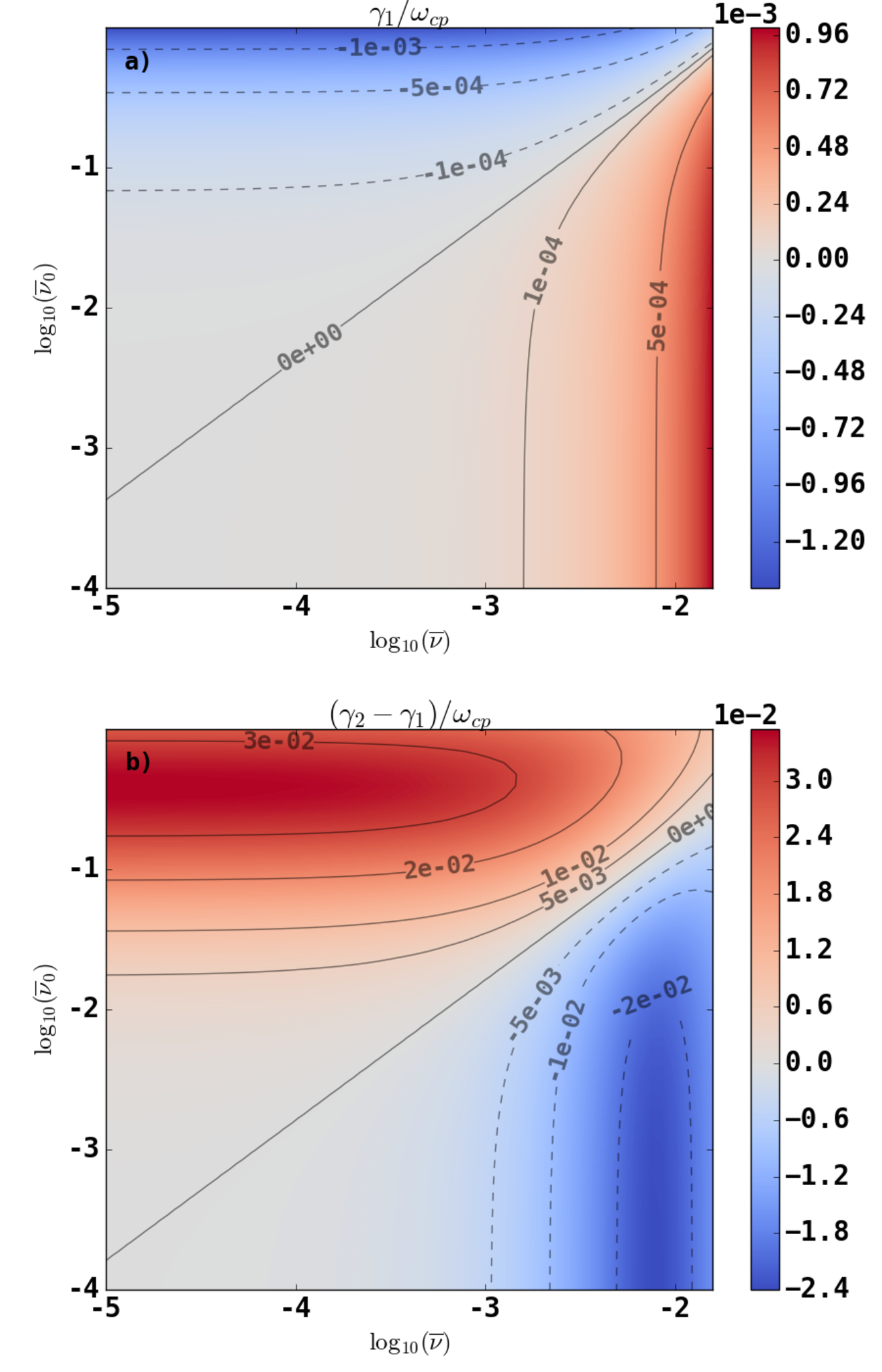}
  \caption{Dependence of $\gamma_1$ (a) and $\gamma_2-\gamma_1$ (b) in the $(\nu,\nu_0)$ parameter space with $Z_1=Z_2=1$, $A_1=4$, $A_2=3$, when the first species is forced so that $\Omega_E=\Ommax{1}$. }
  \label{fig:He}
\end{figure}

The results for a simple case where the forced species is hydrogen and the second species is deuterium is shown in Fig~\ref{fig:HD}. It is seen that depending whether $\nu$ or $\nu_0$ dominates, $\gamma_1$ is positive or negative. However, $\gamma_2-\gamma_1$ is always positive (ejection), as anticipated, and its values are an order of magnitude higher than $\gamma_1$. This is a very favorable case for separation, it means the heavy impurity will be easily ejected radially, compared to the first forced species, which is either confined, or ejected along with the electrons on much slower time scales.

\subsubsection{\label{subsubsec:paramspace_light} Light impurity}

The results of the case where the first species is $^4$He and the second is $^3$He, with $Z_1=Z_2=1$, is shown in Fig~\ref{fig:He}. Again, it confirms our physical intuition formulated at the beginning of this section, namely, that $\gamma_2-\gamma_1$ is positive when the neutrals dominate the friction, and negative when the Coulombian collisions dominate. 

\subsection{\label{subsec:trace}Conclusion on the case of trace impurity}

As far as we could tell from our numerical experiments, the qualitative patterns observed in Figs.~\ref{fig:HD} and \ref{fig:He} is quite general. The parameter that matters is whether the second impurity, that which does not undergo direct azimuthal forcing, is heavier (figure~\ref{fig:HD}) or lighter (figure~\ref{fig:He}).

This is an important new result of this work. Indeed, conventional wisdom of the Ohkawa filter suggests that the radially ejected species is always the heavy one. Our results challenge this idea. When the friction with neutrals dominates over the Coulombian collisions, a situation that is common for this type of plasma\cite{gueroultNecessaryConditionPerpendicular2019}, a light impurity can be expelled radially. This remains true even if the two species have masses that are close to each other, \emph{e.g.} isotopes of heavy elements. 

Note that even when $\gamma_1>0$, the latter remains an order of magnitude lower than $\gamma_2$. This confirms the potential for isotopic separation. This conclusion relies crucially on the hypothesis that only one of the two species (the one that is supposed to be confined) undergoes the azimuthal forcing which is the source of the electric field. Note that increasing $\nu_0$, the friction with neutrals, while keeping $\nu$ low, tends to have a beneficial impact on the separation capability. However, one should keep in mind that this leads to an increase of $F$, and therefore to an increase of the power consumption of the device. At the same time, since the separation time scale becomes shorter, higher fluxes might be achieved to compensate. It is not obvious that the favorable parameters for extraction could be easily achieved. Increasing friction with the neutrals can be done by increasing their density, while decreasing $\nu$ can be done by raising the temperature. However, the two are contradictory, or at least, counter-intuitive, since increasing the temperature shifts the ionization equilibrium of Saha law toward less neutrals. A more thorough investigation considering the time scales of ionization, injection and particle residence time in the device would be needed to conclude. However, the considerations of ion production mechanisms are well beyond the scope of the present work.


\section{\label{sec:notrace}Finite concentration}

There remains to examine the important case of finite concentration. In general, the interesting samples to separate may have two species with comparable abundances, or even a large number of species. We will consider only the two species case and examine a last mathematical possibility offered by the model.

When the plasma consists of two species with comparable densities and different charge to mass ratios, there are now two values for the Brillouin limit, $\Ommax{1}\neq\Ommax{2}$, and it is difficult to have the correct intuition about how the plasma is going to react to the forcing. Once again, we distinguish the cases of heavy and light impurity.

\subsection{\label{subsec:finiteheavy}Heavy second species}

We still assume that the first species is forced, while the second is not. Let us consider first the case when the second species is heavy, with $\Ommax{1}>\Ommax{2}$. If the radial velocities start off as in the case of a trace heavy impurity, that is with $\gamma_2$ positive and $|\gamma_2|>|\gamma_1|$ whatever the sign of $\gamma_1$, the outflow of the second species leads to a rapid decrease of its density, leaving place to only the first ion species and the electrons. We are then back to the situation of a trace species. The situation, then, is essentially non stationary, and it does not make sense to fix $\eta$ and $\zeta$ (parameterizing the initial respective abundances of the two species) and ask what are the stationary electric field and frequencies for these values of $\eta$ and $\zeta$, given a forcing $F$ as well as the friction frequencies $\nu$ and $\nu_0$. Indeed, $\eta$ and $\zeta$ will necessarily vary, and in fact $\eta$ will go exponentially to zero. Unfortunately, we have not come up with a better way than to actually simulate the evolution of the system for a range of parameters and examine the final state. This is unfortunate because the numerical time required to reach a steady state increases roughly inversely proportionally to the collision frequency. We explore again the parameter space, with the following algorithm. For a given choice of $A_1$, $A_2$, $Z_1$ and $Z_2$, we fix $\nu$ and $\nu_0$ and choose the forcing such that $\Omega_E$ would be equal to $\Ommax{1}$ if the second species was a trace impurity. Finally, we choose initial values for $\eta_0\equiv\eta(t=0)$ between 0 and 1 (and $\zeta_0=1-\eta_0$) and we let the coupled system evolve according to the full dynamical equations and wait until a stationary state is reached. When we do the exercise for $A_1=1$ and $A_2=2$ (this is figure~\ref{fig:HD} for the case of trace impurity), testing the values $\eta_0$=0, 0.2, 0.4, 0.6 and 0.8, we obtain that for all the possible initial values without exception, the final state has $\eta\to0$.

The conclusion is that at least in the case of $A_1=1$ and $A_2=2$, there is no difference between the finite concentration case and the trace impurity case. The result is that the second species is radially ejected. Of course, we cannot prove that this holds for arbitrary masses such that $A_2>A_1$, but the model provides a way to check relatively rapidly whether this is true or not, for any particular case one might be interested in.

\subsection{\label{subsec:finitelight}Light second species}
\begin{figure}[t!]
  \centering
  \includegraphics[scale=0.4]{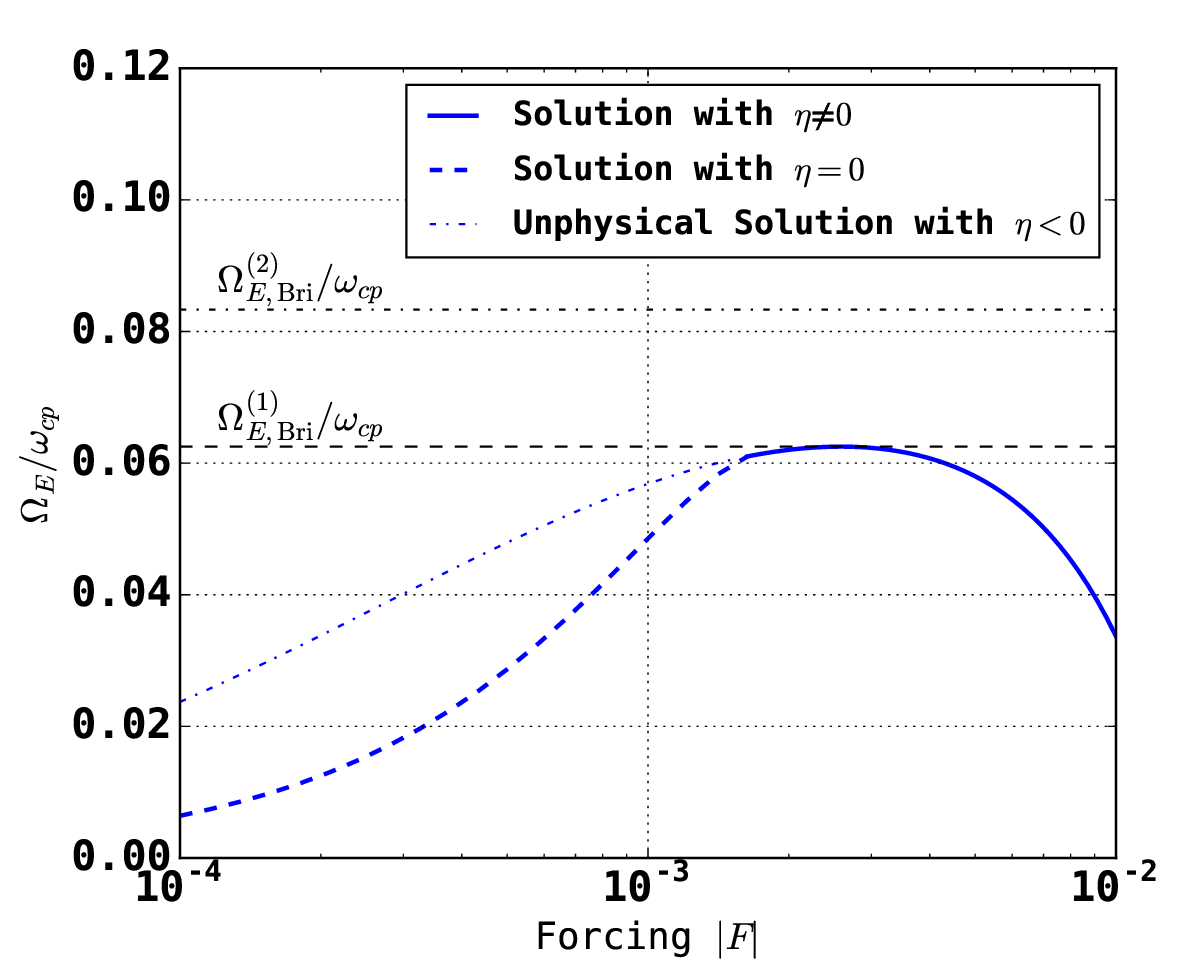}
  \caption{Dependence of $\Omega_E$ with the forcing for $A_1=4$, $A_2=3$, $Z_1=Z_2=1$, $\overline{\nu}=\sci{3}{-3}$ and $\overline{\nu}_0=\sci{3}{-2}$. The solid line represents the solution of \refeq{sys1}-\refeq{sys2}. For $|F|<\sci{1.3}{-3}$ (dash-dotted line), the solution of the system yields negative values for $\eta$. The equilibrium solution then actually has $\eta=0$ and is numerically obtained by simulating the dynamical evolution until stationary state is reached. The case $\eta=0$ corresponds to the previously studied case of trace impurity, and the solution no longer respects the constraint \refeq{gam12e}.}
  \label{fig:gam12e}
\end{figure}

The case of the light species is, again, more interesting. We have seen in the trace impurity case that although $\gamma_2>0$ when neutral friction dominates, the opposite holds, $\gamma_2<0$, when Coulombian friction dominates. In the latter case, the negativity of $\gamma_2$ means an exponential increase of the density. If we are speaking of a trace impurity, this makes no sense: where would the required matter come from? However, this hints at a last mathematical possibility, which we have not explored yet. What if there was a stationary state with $\eta$ and $\zeta$ both non zero?

If we seek such a stationary state $\eta\ne0$, then we are bound to have
\begin{align}
  \label{eq:gam12e}
\gamma_1=\gamma_2=\gamma_e  
\end{align}
with
\begin{align}
  \label{eq:gamestationary}
  \gamma_e = -\abrack{\overline{\nu}\zeta Z_1\parent{\Omega_1+\Omega_E} +\overline{\nu}\eta Z_2\parent{\Omega_2+\Omega_E} +\overline{\nu}_0\sqrt{\frac{m_e}{m_p}}\Omega_E},
\end{align}
where $\Omega_E$ is a function of $\zeta$ and $\eta$ \emph{via} \refeq{efield}. We see that indeed $\dot{\Omega}_E$ from Equation \refeq{efielddot} vanishes when \refeq{gam12e} holds. Equation \refeq{efielddot} may have a different stationary equilibrium for some value of $\zeta$ and $\eta$, but then equations~\refeq{zetadot}-\refeq{etadot} would lead to a change of densities, and the stationary solution for $\Omega_E$ would not be preserved. We then have, with equations \refeq{gammadot}-\refeq{Omegadot} for the two species (without the forcing for the second one), a system of four equations for the four remaining unknowns $\Omega_1$, $\Omega_2$, $\eta$ and $\zeta$. In principle, we can look for a stationary point for these variables. According to equation~\refeq{gam12e}, all species are confined or expelled with the same radial velocities, which is not good for separation. The question, then, is that of the existence and stability properties of that fixed point, from a dynamical system point of view. 

Let us examine the fixed point given by the condition \refeq{gam12e}. In the general case where $\nu$ and $\nu_0$ are both nonzero, since at equilibrium the radial Coulombian friction terms vanish because of \refeq{gam12e}, the system reads
\begin{widetext}
  \begin{align}
    \label{eq:sys1}
  &\Omega_1^2-\gamma_e^2 + \omega_{c1}\parent{\Omega_1+\Omega_E} - \frac{1}{\sqrt{A_1}}\nu_0\gamma_e=0\\
    \label{eq:sys2}
  &-\parent{2\Omega_1+\omega_{c1}}\gamma_e + \omnorm^2F - \nu_{12}\parent{\Omega_1-\Omega_2}-\frac{Z_1^2}{A_1}\nu\parent{\Omega_1+\Omega_E} - \frac{1}{\sqrt{A_1}}\nu_0\Omega_1=0\\
    \label{eq:sys3}
  &\Omega_2^2-\gamma_e^2 + \omega_{c2}\parent{\Omega_2+\Omega_E} - \frac{1}{\sqrt{A_2}}\nu_0\gamma_e=0\\
    \label{eq:sys4}
  &-\parent{2\Omega_2+\omega_{c2}}\gamma_e - \nu_{21}\parent{\Omega_2-\Omega_1}-\frac{Z_2^2}{A_2}\nu\parent{\Omega_2+\Omega_E} - \frac{1}{\sqrt{A_2}}\nu_0\Omega_2=0
\end{align}  
\end{widetext}
with $\gamma_e$ given by \refeq{gamestationary} and $\Omega_E$ given by \refeq{efield}. The solutions of \refeq{sys1}-\refeq{sys4} are not restricted, mathematically, to $\zeta,\eta>0$, although this is a physical requirement. Therefore, we consider there is no physical solution satisfying \refeq{gam12e} when the solution of \refeq{sys1}-\refeq{sys4} is such that $\eta$ or $\zeta$ is negative. It is difficult to prove the existence or non existence of relevant solutions in the general case. But we can still say an important thing on the solutions. As we have repeatedly seen, the equilibrium in the radial direction is dominated by the equilibrium between inertial, Lorentz and electric forces, $\Omega_j^2+\omega_{cj}(\Omega_j+\Omega_E)=0$ for any ion species $j$, which allows to define the Brillouin limit. If $\Omega_E$ ever overcomes the Brillouin limit, the other component of the inertial forces, $\gamma_j^2$, is bound to come into play and assume a large value, corresponding to the radial ejection of the species. Therefore, a stationary state such that $\Ommax{1}<\Omega_E<\Ommax{2}$ would require $\gamma_1$ large but $\gamma_2$ small. Since the equilibrium we are considering has $\gamma_j=\gamma_e\,\forall j$, by \emph{reductio ad absurdum}, we can conclude that $\Omega_E$ can approach $\Ommax{1}$ (which is the minimum of the two limits since we assume the second species is lighter) but never exceed it. This is illustrated by figure~\ref{fig:gam12e}. Therefore, the equilibria we are considering here never have large radial ejection velocities, although friction can lead to $\gamma_e>0$ as we have seen.

We can do an exercise similar to figure~\ref{fig:He} where we look for a solution of the type~\refeq{gam12e} in the $(\nu,\nu_0)$ parameter space, assuming the same rule for $F$ as before (we take for $F$ the value that would make $\Omega_E$ maximum if the second species were a trace). For some values of $(\nu,\nu_0)$ we find physically acceptable solutions where both $\eta$ and $\zeta$ are positive, but for others either $\zeta$ or $\eta$ is negative. The boundary between the two regions coincides precisely with the isocontour $\gamma_2-\gamma_1=0$, visible in figure~\ref{fig:He} (we do not show the result as the figure would not bring any new information). This coincidence can be understood using the following argument. Let us fix $\nu$, $\nu_0$ and $F$ and examine the dynamical picture, starting from nonzero values for $\eta$ and $\zeta$ and $\Omega_E=0$ at $t=0$. By linearizing the system, it can be proven that for early times, we have $\gamma_2>0$ and $\gamma_1<0$, so the first species is confined and the second is expelled. First, assume that $\nu$ and $\nu_0$ are such that in the trace case $\gamma_2-\gamma_1<0$. As the second species is expelled, its density exponentially decreases until it becomes a trace, but by assumption when it has become a trace, $\gamma_2-\gamma_1<0$ and we are now in the opposite situation where the expelled species is the first one. This hints at the possibility of an equilibrium point with $\eta\ne 0$. On the contrary, when in the trace case $\gamma_2-\gamma_1>0$, there is nothing to stop the ejection process of the second species and its density can decrease exponentially until it becomes zero. 

Therefore, the solutions conforming to \refeq{gam12e} can be found only in the zone where $\gamma_2-\gamma_1<0$ of the trace case. Referring to figure~\ref{fig:He}, this is the bottom-right part of the $\gamma_2-\gamma_1$ figure, colored in blue.

\section{\label{sec:Discussion}Discussion and caveats}

The results of this paper can be reformulated in practical terms, if one is interested in the physical parameters of an actual experiments. First one should have the plasma sufficiently magnetized such that the Coulomb collision frequency is small, $\nu<\omega_{cp}$. This sets a minimum value to the magnetic field, or to the plasma temperature, or a maximum value to the electron density. Then, if one wants to collect along the stream a light species, and eject the other heavy one, it is sufficient to tune the wave to be resonant with the light species, which will be confined, while the heavy species will be expelled radially. If one wants to collect along the stream the heavy species and eject radially the light one, it is also possible, but it requires that the collisions be dominated by collisions with neutrals, that is $\nu\ll\nu_0$. Increasing $\nu_0$ can be done by increasing the neutral density, for instance by decreasing the temperature, but then the question is whether these neutrals are from the stream one wishes to separate. If the neutral fraction is very large, then the device loses its purpose. Also, if the neutral density is so large that collisions significantly impede rotation and electric field build up, the electrical consumption of the device will increase. It is beyond the scope of this paper to discuss these problems in more details.

We also have to discuss the limits of the present study. Our results come with important caveats, that have been already mentioned or have transpired throughout the exposition. 

First of all, we never make explicit the coupling between the wave and the plasma. Actually, before even considering the resonant wave-particle interaction, the wave propagation should be considered, as several interesting effects due to both the rotation and the magnetic field occur. In fact, it appears that the dispersion relation of an orbital angular momentum carrying wave in a rotating magnetized plasma is not known. It is known that in a rotating magnetized plasma, there is, in addition to the Faraday birefringence that causes a rotation of the polarization plane, a mechanical optical effect, where rotation also contributes to the rotation of the polarization plane\cite{gueroultDeterminingRotationDirection2019}. However, the dispersion relation is known formally only when the wave number $\K$, $\bm{\Omega}$ and $\B_0$ are aligned. For oblique propagation, including the case of wave angular momentum, there are additional terms in the wave equation, that are not straightforward to treat analytically because of the tensorial nature of the susceptibility in a magnetized plasma. Therefore, it is a relevant question whether \emph{e.g.} the conditions of resonance at $\Omega_1=0$ would be preserved when $\Omega_1$ takes on values of the order of the ion cyclotron frequency. Translated in more practical terms, would $F$ depend on $\Omega_1$?

We regard the problem of the shaping of power deposition as less pressing. If power deposition is very different from our assumption that $\mathcal{P}(r)\propto r^4$, this may introduce shear and shaping of the electric field, but not necessarily invalidate our results. More studies would be required to conclude, but we would lose the immense advantage of dealing with ODEs.

As mentioned in section~\ref{subsec:Notations}, we have not considered the influence of the potentially large poloidal currents associated with the ion-electron slip. This slip is of order 1 in our normalization. Using Ampère's theorem on a section of cylinder of length $L$ and using $\upsilon_\theta\propto r$, integrating up to the cylinder radius $r$, we find a $\delta B$ perturbation of the order of
\begin{align}
  \frac{\delta B}{B_0} &= \mu_0\frac{a^2}{2}n_ee\frac{\omega_{cp}}{B_0}\parent{\zeta Z_1\Omega_1-\Omega_e}\\
                       &=\frac{1}{2}\frac{n_ee^2}{\epsilon_0 m_p}\frac{a^2}{c^2}\parent{\zeta Z_1\Omega_1-\Omega_e}
\end{align}
(assuming a one species plasma to simplify). Obviously, this perturbation can become large unless $a$ is less than the skin depth $c/\omega_{pi}$. This is a quite severe restriction on the range of plasma parameters than can be achieved. The question of how the dynamics would be affected if this perturbation was taken into account is left for future work. Note that some insight to this issue is provided by ref.~\cite{gueroultTendencyRotatingElectron2013}

Another limitation of this work, also pertaining to the 0-D nature of the model, has already been discussed. We have completely neglected the boundary conditions, both in the $z$ direction and at the cylinder walls in $r=a$. The neglect of boundary conditions in the $z$ direction is relatively natural since it amounts to assume a long cylinder, but the issue of the radial boundary conditions is more problematic. As a result of not considering these boundary conditions, there is no limitation on the variations of density that can be achieved by the model. It is as if there were perfect reservoirs or sinks in $r=a$, or, alternatively, as if $a$ were in effect infinite. We have been careful in this paper not to get carried away when the model's behaviour was clearly caused by taping in these reservoirs. For instance, when a trace impurity is sucked until it constitutes the main plasma component, while the initially dominant species is expelled, we consider this conclusion to be problematic. We believe, however, that we have been careful enough with this problem that the main conclusions of the paper, summarized in section~\ref{sec:Summary}, still hold. More advanced models should take this problem explicitly into account, by considering global conservation of the number of particles.

\section{\label{sec:Summary}Summary}

We can now summarize the results of this paper.
\begin{itemize}
\item The most important contribution of this paper is the realization that when the source of the electric field is azimuthal forcing, the Brillouin limit with collisions cannot be understood independently of the forcing. When this is accounted for, we find there is no breakdown of the Brillouin limit due to collisions (section~\ref{sec:BrillouinBreakdown}), contrary to the case where the source of the electric field is biased electrodes~\cite{raxBreakdownBrillouinLimit2015}. This essential difference between the two different ways to generate the electric field appears to have been overlooked in the previous literature on the Ohkawa filter.
\item In the case of a one species plasma, the Brillouin limit decreases for large values of the friction with neutrals, and is not affected by Coulombian friction unless the collision frequency reaches irrelevant values for the plasmas we consider (sections~\ref{subsec:ColBrilNeutral} and~\ref{subsec:ColBrilCharged}).
\item When we consider trace impurities, assuming the forcing to be such that $\Omega_E=\Ommax{1}$, we find that
  \begin{enumerate}[label=(\roman*)]
  \item The trace impurity is always expelled when it is heavier (section~\ref{subsubsec:paramspace_heavy}).
  \item The trace impurity can be expelled when it is lighter if the friction is dominated by neutrals, which, by the way, is likely (section \ref{subsubsec:paramspace_light}).
  \end{enumerate}
\item When the plasma consists of two species with comparable abundances, with the same assumption on the forcing, we again have two cases:
  \begin{enumerate}[label=(\roman*)]
  \item If the passive species is the heavy one, it is always expelled (section \ref{subsec:finiteheavy}).
  \item If the passive species is the light one, the dynamical system can exhibit a fixed point with $\eta$ and $\zeta$ both non-zero, and $\gamma_1=\gamma_2=\gamma_e$, if the friction is dominated by Coulomb collisions. Otherwise, it is expelled (section \ref{subsec:finitelight}).
  \end{enumerate}
\item The main limits of the present results are as follows:
  \begin{enumerate}[label=(\roman*)]
  \item The radius of the plasma must be smaller than the ion skin depth to justify neglect of the magnetic field perturbation.
  \item The coulping between the wave and the plasma requires specific studies.
  \item The boundary conditions, that is, sources and sinks, are not investigated in any detail.
    \end{enumerate}
  \end{itemize}




 
\section*{Data availability}

The data that support the findings of this study are available from the corresponding author upon reasonable request.

\begin{acknowledgments}
  The author would like to thank X. Garbet for insightful discussion and encouragements.
\end{acknowledgments}

\appendix

\section{\label{ap:justif} Justification for the force term}

\begin{figure}[h!]
  \centering
  \includegraphics[scale=0.6]{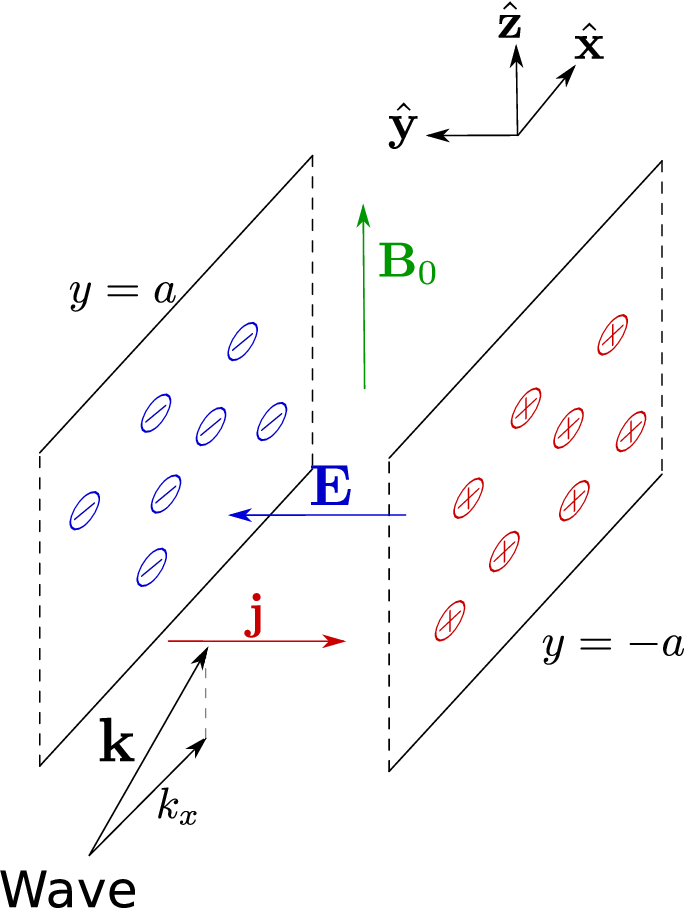}
  \caption{Configuration with wave induced electric field}
  \label{fig:mom}
\end{figure}

We will analyze the configuration represented in figure \ref{fig:mom}. The plasma is infinite in the $\xhat$ direction (and possibly in the ignorable $\zhat$ direction as well). The plasma is bounded in the $\yhat$ direction, in $y=\pm a$. There is resonant interaction between a wave with wave vector $\K_\perp=k_\perp\xhat$ and the plasma. We will adopt two different points of view on the effect of this wave on the plasma. The two points of view reach the same conclusion: the momentum of the wave is entirely converted to plasma and dc field momentum, with proportions respectively $\chi_\perp/\epsilon_\perp$ and $1/\epsilon_\perp$. 

In the following, momentum refers to the kinetic part only, ${\bf p} = m\V$, rather than to the canonical momentum ${\bf p}=m\V+q{\bf A}$.

\subsection{\label{subap:particle}Particle point of view}
Let's assume that during a time $\Delta t$, each particle of a minority resonant population with density $n_r$ absorbs momentum $\Delta {\bf p}=\Delta p\xhat$, with $\Delta p\propto\hbar k_\perp>0$. The total momentum absorbed per unit time and unit volume, that is, the density of momentum transfer between the wave and the plasma, is

\begin{align}
  \label{eq:F}
  {\bf F} = n_r\frac{\Delta{\bf p}}{\Delta t}.
\end{align}

The gyrocenter of the resonant particles is displaced by a quantity
\begin{align}
  \Delta y = -\frac{\Delta p}{qB_0},
\end{align} which derives simply from the expression of the Larmor radius.
This corresponds to a current of free charges $\J_r = j_r\yhat$, where
\begin{align}
  j_r &= n_rq\frac{\Delta y}{\Delta t}\nonumber\\
    &=-n_r\frac{\Delta p}{B_0\Delta t}
\end{align}

This current of free charges accumulates positive charges on the right, $y=-a$, while depleting charges in $y=a$. Surface charges appear at the left and right boundaries, given by $\Delta \sigma = j_r\Delta t$. However, the plasma is a dielectric with susceptibility $\chi_\perp = mn/\epsilon_0B^2$, and dielectric constant $\epsilon_\perp = 1+\chi_\perp$. Although in the plasmas we're interested in, $\chi_\perp\gg1$, we will not use this assumption and maintain the distinction between $\epsilon_\perp$ and $\chi_\perp$. Otherwise we might lose track of some momentum. Since the plasma is a dielectric, the free charge current $j_r$ is largely compensated by a bulk polarization current in the opposite direction. The electric field increase is therefore given by
\begin{align}
  \label{eq:efield_ap}
  \epsilon_0\epsilon_\perp\Delta \E &= -j_r\Delta t\yhat\nonumber\\
                                    &=n_r\frac{\Delta p}{B_0}\yhat
\end{align}


The associated cross-field motion is
\begin{align}
  \Delta \V_E = \frac{\Delta E}{B_0}\xhat.
\end{align}

The momentum density of the electromagnetic field is given by $\textsf{\textbf{P}}_\mathrm{field} = \epsilon_0\E\times\B$. Here, we do not bother with the Abraham-Minkowski controversy, which debates whether the relevant electromagnetic momentum in matter is ${\bf D}\times\B$ (Minkowski) or $\E\times{\bf H}/c^2$ (Abraham). After all, the quantities ${\bf D}$ and ${\bf H}$ are but useful auxiliary quantities, which allow to hide the polarization charge and magnetization currents. If we write all the currents explicitly, we are authorized to stick with the usual expressions for field quantities. Therefore, in the final situation, the field momentum increase per unit volume is
\begin{align}
  \Delta \textsf{\textbf{P}}_\mathrm{field} &= \epsilon_0\Delta E B_0\xhat\nonumber\\
                                            &=\frac{n_r\Delta p}{\epsilon_\perp}\xhat,
\end{align}
while the increase of mechanical momentum is
\begin{align}
  \Delta \textsf{\textbf{P}}_\mathrm{mech} &= mn\Delta\V_E \nonumber\\
                                           &= \frac{\chi_\perp}{\epsilon_\perp}n_r\Delta p \xhat
\end{align}
And we find that
\begin{align}
  \Delta\parent{\textsf{\textbf{P}}_\mathrm{field} + \textsf{\textbf{P}}_\mathrm{mech}} = n_r\Delta p \xhat
\end{align}

The momentum transferred by the wave goes to the plasma and the dc field in proportions respectively $\chi_\perp/\epsilon_\perp$ and $1/\epsilon_\perp$.  Incidentally, note that ${\bf D}\times\B$ gives the sum of the field and matter contributions.


\subsection{\label{subap:fluid}Fluid point of view}

Now let's have a look at the fluid point of view. We only need two equations. The electron dynamics is useless because of their small inertia. To study the ion motion under the influence of the wave, let us start from the kinetic equation taking into account the quasilinear wave-induced velocity diffusion, and neglecting collisions, as well as spatial gradients:

\begin{align}
  \partial_t f + \frac{q}{m}\parent{\E+\V\times\B}\cdot\partial_\V f = \partial_\V\cdot\cbrack{\bar{\bar{{\bf D}}}\cdot\partial_\V f}
\end{align}

By integrating on velocity space after multiplying by $m\V$, and using constant density, we find the equation of motion of the ions:

\begin{align}
  \partial_t\V = \frac{q}{m}\parent{\E+\V\times\B} + \frac{\F}{mn},
\end{align}
where $\F$ is the wave-induced forcing, defined by
\begin{align}
  \F = m\int\V\partial_\V\cdot\cbrack{\bar{\bar{{\bf D}}}\cdot\partial_\V f}\ud^3\V  = -m\int\ud^3\V \bar{\bar{{\bf D}}}\cdot\partial_\V f
\end{align}
By conservation of momentum, we know that $\F$ should be equal to the momentum lost by the wave, hence $\F$ is also given by equation \refeq{F}:
\begin{align}
  \F = n_r\frac{\Delta p}{\Delta t}\xhat
\end{align}

The dynamics of the electric field $\E = E\yhat$ is given by the same equation as in section~\ref{subap:particle} with the total current ${\bf j} = nq\V$:
\begin{align}
  \epsilon_0\partial_t E = -nq\upsilon_y
\end{align}

Since there is no spatial dependence except for the charge accumulation at the $y$ boundaries, we can transform the system into a system of coupled ODEs, ignoring the $z$ direction. Writing $\upsilon_E = E/B_0$, the system reads

\begin{align}
  &\dot{\upsilon}_x = \omega_c \upsilon_y + \omega_c\mathcal{F}\\
  &\dot{\upsilon}_y = -\omega_c \upsilon_x + \omega_c\upsilon_E\\
  &\dot{\upsilon}_E = -\frac{nm}{\epsilon_0 B_0^2}\omega_c \upsilon_y = -\chi_\perp \omega_c \upsilon_y.
\end{align}
where $\mathcal{F} = |{\bf F}|/(nqB_0)$ represents the wave forcing. We could already stop here. Indeed, the total momentum (neglecting the $q{\bf A}$ contribution) rate of change is
\begin{align}
  \partial_t\cbrack{\textsf{\textbf{P}}_\mathrm{field} + \textsf{\textbf{P}}_\mathrm{mech}} &= mn \dot{\upsilon}_x + \epsilon_0\dot{E}B_0\nonumber\\
                                                                                            &= mn\parent{\dot{\upsilon}_x + \dot{\upsilon}_E/\chi_\perp}\nonumber\\
                                                                                            &=\omega_c\mathcal{F},
\end{align}
and we know the proportions (the same as in section~\ref{subap:particle}, respectively $\chi_\perp/\epsilon_\perp$ and $1/\epsilon_\perp$), because the $\upsilon_x$ motion is nothing but the $\E\times\B$ motion. But let's compute the analytical solution to see in details what happens. After normalizing time to $\omega_c^{-1}$, and defining $X = (\upsilon_x,\upsilon_y,\upsilon_E)^\top$, the system reads
\begin{align}
  \dot{X} = AX + B
\end{align}
with
\begin{align}
  A =
  \left(
  \begin{array}{ccc}
    0&1&0\\
    -1&0&1\\
    0&-\chi_\perp&0
  \end{array}
             \right),
             \qquad
             B = \left(
             \begin{array}{c}
               \mathcal{F}\\
               0\\
               0
             \end{array}
  \right)
\end{align}

The solution satisfying $X(t=0)=0$ is given by

\begin{align}
  X(t) = M(t)B,\qquad M(t) = \parent{\sum_{n=1}^\infty \frac{A^{n-1}t^n}{n!}}
\end{align}

We can easily find the analytical solution by noting the following property of the matrix $A$:
\begin{align}
  &A^{2k} = (-1)^{k-1}\epsilon_\perp^{k-1}A^2\\
  &A^{2k-1} = (-1)^{k-1}\epsilon_\perp^{k-1}A,
\end{align}
with
\begin{align}
  A^2 =   \left(
  \begin{array}{ccc}
    -1&0&1\\
    0&-\epsilon_\perp&0\\
    \chi_\perp&0&-\chi_\perp
  \end{array}
             \right)
\end{align}

The matrix $M(t)$ is then given by
\begin{align}
  M(t) = t1 + \frac{A}{(i\sqrt{\epsilon_\perp})^2}\sum_{k=1}^\infty\frac{(it\sqrt{\epsilon_\perp})^{2k}}{(2k)!} + \frac{A^2}{(i\sqrt{\epsilon_\perp})^3}\sum_{k=1}^\infty\frac{(it\sqrt{\epsilon_\perp})^{2k+1}}{(2k+1)!}
\end{align}
We recognize in the two sums truncated developments of sine and cosine, which yields finally:
\begin{align}
  &\upsilon_x(t) = \mathcal{F}\parent{\frac{\chi_\perp}{\epsilon_\perp}\omega_ct + \frac{1}{\epsilon_\perp^{3/2}}\sin\parent{\epsilon_\perp^{1/2}\omega_c t}}\\
  &\upsilon_y(t) = \frac{\mathcal{F}}{\epsilon_\perp}\parent{\cos\parent{\epsilon_\perp^{1/2}\omega_c t} - 1}\\
  \label{eq:vEsol}
  &\upsilon_E(t) = \mathcal{F}\parent{\frac{\chi_\perp}{\epsilon_\perp}\omega_ct - \frac{\chi_\perp}{\epsilon_\perp^{3/2}}\sin\parent{\epsilon_\perp^{1/2}\omega_c t}}
\end{align}

In other words, up to the oscillating terms, that become small anyway when $\chi_\perp\to\infty$, we have exactly the same dynamics as in section~\ref{subap:particle}, and the same conclusion. Indeed, in a time $\Delta t$, the combination plasma + dc field receives the momentum

\begin{align}
  mn(\upsilon_x(\Delta t)+\upsilon_E(\Delta t)/\chi_\perp) = mn\mathcal{F}\omega_c\Delta t = n_r\Delta p
\end{align}

The momentum of the wave is transferred to the plasma and the dc field in proportions respectively $\chi_\perp/\epsilon_\perp$ and $1/\epsilon_\perp$.

\subsection{\label{subap:splitting}Splitting between resonant and bulk}

One might object that a fluid approach is bound to fail because the population that undergoes wave forcing, the resonant particles, makes up such a small fraction of the plasma. In this section, we split the ions between a resonant part, with density $n_r=\alpha n$, and a bulk part, with density $n_b = (1-\alpha)n$. Only the resonant part sees the forcing ${\bf F}$. Note that the distribution of the resonant particles is anything but a Maxwellian in the $z$ direction, but can be close to a Maxwellian (insofar as the bulk also is) in the $xOy$ plane, because there is no correlation between having a certain velocity (close to the phase velocity of the wave in the $z$ direction) and the velocity in the $xOy$ plane. Hence, we can make sense of treating the resonant part as a fluid in the $xOy$ plane. 

We show that the dynamics of the electric field (hence, of the bulk motion) is completely independent of $\alpha$. Therefore, there is no reason to split resonant ions and bulk ions, and the approach where ${\bf F}=\K_\perp P/\omega$ is seen as a fluid force is valid. This crucially relies on the fact we are examining the perpendicular direction.

One would like to solve the following coupled equations (the convective term is still absent because of the geometry of the slab system and the direction of the flows):

\begin{align}
  \label{eq:SplitSystem_vr}
  n_rm\partial_t\V_r &= n_rq\parent{\E+\V_r\times\B} + \F\\
  \label{eq:SplitSystem_vb}
  n_bm\partial_t\V_b &= n_bq\parent{\E+\V_b\times\B} \\
  \label{eq:SplitSystem_E}
  \epsilon_0\partial_t E &= -(n_r+n_b)q\parent{\frac{n_r}{n_r+n_b}v_{r,y}+ \frac{n_b}{n_r+n_b}v_{b,y}}\nonumber\\
                     &=-nq\parent{\alpha v_{r,y}+ (1-\alpha)v_{b,y}},
\end{align}
where we have defined $\alpha\equiv n_r/(n_r+n_b)$.
As before, we normalize the system, in particular time with $\omega_{ci}$, defining $\chi_\perp = (n_r+n_b)m/(\epsilon_0 B^2)$, $v_E=E/B$, and $\mathcal{F} = |\F|/(nqB_0)$. We obtain the $5^\mathrm{th}$ order system

\begin{align}
  \label{eq:splitsystemnorm}
  &\dot{v}_x = v_y + \frac{\mathcal{F}}{\alpha}\\
  &\dot{v}_y = v_E-v_x \\
  &\dot{w}_x = w_y\\  
  &\dot{w}_y = v_E - w_x\\
  \label{eq:vEsplit}
  &\dot{v}_E = -\chi_\perp\parent{\alpha v_y + (1-\alpha)w_y},
\end{align}
where the $v$ variables stand for the resonant part, and the $w$ variables for the bulk.

The important result is that the dynamics of the electric field and of the transverse current (in the $y$ direction) do not depend on $\alpha$: whatever the value of $\alpha>0$, we obtain the same electric field evolution, and the same mean displacement of the ions in the $y$ direction (representing the radial direction of a cylindrical system). We can prove it as follows. By taking the time derivative of (\ref{eq:vEsplit}), one obtains
\begin{align}
  \ddot{v}_E + \chi_\perp v_E = \chi_\perp\parent{\alpha v_x + (1-\alpha) w_x}
\end{align}
By taking it one more time and using the notation $\mathcal{Y}\equiv \dot{v}_E$, one obtains
\begin{align}
  \ddot{\mathcal{Y}} + \chi_\perp\mathcal{Y} = \chi_\perp\parent{\alpha\abrack{v_y+\frac{\mathcal{F}}{\alpha}} + (1-\alpha)w_y},
\end{align}
which, using $\epsilon_\perp=1+\chi_\perp$ and, one more time, equation (\ref{eq:vEsplit}), simplifies into
\begin{align}
  \label{eq:Y}
  \ddot{\mathcal{Y}} + \epsilon_\perp\mathcal{Y} = \chi_\perp \mathcal{F}.
\end{align}
The parameter $\alpha$ has disappeared from the equation. The evolution of $E$, therefore, is independent of $\alpha$, with the condition that the 3 initial conditions of this third order system also be independent of $\alpha$. Rewriting the system $\dot{X} = AX + B$, with $X=(v_x,v_y,w_x,w_y,v_E)^\intercal$, $B=(\mathcal{F}/\alpha,0,0,0,0)^\intercal$, and
\begin{align}
  \label{eq:Amat}
  A = \left(
  \begin{array}{ccccc}
    0&1&0&0&0\\
    -1&0&0&0&1\\
    0&0&0&1&0\\
    0&0&-1&0&1\\
    0&-\alpha\chi_\perp&0&-(1-\alpha)\chi_\perp&0
  \end{array}
  \right),
\end{align}
the solution is
\begin{align}
  \label{eq:Xt}
  X(t) &= \abrack{\sum_{n=1}^\infty \frac{A^{n-1}t^n}{n!}}B\nonumber\\
       &=B t + AB\frac{t^2}{2}+\mathcal{O}(t^3)
\end{align}
Therefore, the initial conditions for $v_E$ are $v_E(0)=0$, $\dot{v}_E(0) = 0$ and $\ddot{v}_E(0) =0$, also independent of $\alpha$. Since the total $y$-current of this species is simply given by the time derivative of $v_E$, this current also is independent of $\alpha$. QED.

With these initial conditions, we can actually easily solve for $v_E$. From equation (\ref{eq:Y}), we have the solution
\begin{align}
  \mathcal{Y}(t) = \frac{\chi_\perp}{\epsilon_\perp}\mathcal{F}\abrack{1 - \cos\parent{\sqrt{\epsilon_\perp}t}},
\end{align}
which, naturally, gives for $v_E$ the very same solution as equation (\ref{eq:vEsol}).

The conclusion is that there is no point in splitting the fluid between resonant and non-resonant part to treat the wave forcing, whether in terms of electric field dynamics or in terms of net displacement in the $y$ (``radial'') direction. Of course, $\V_r$ and $\V_b$ are different. In particular, when the wave has momentum in the $+\xhat$ direction, the resonant ions are displaced on average to the $-\yhat$ direction, while the bulk ions are displaced to the $+\yhat$ direction. But when one is doing a separation experiment, one is not interested to know whether the collected ions are the resonant ones or not. One is only interested in the species of the ion.

\end{document}